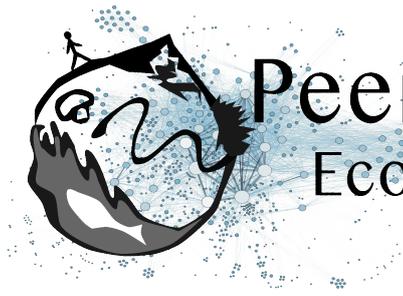

# Peer Community In Ecology

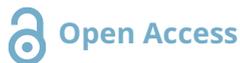
Open Access

RESEARCH ARTICLE

# Impact of group management and transfer on individual sociality in Highland cattle (*Bos Taurus*)

Sebastian Sosa, Marie Pelé, Elise Debergue, Cédric Kuntz, Blandine Keller, Florian Robic, Flora Siegwalt-Baudin, Camille Richer, Amandine Ramos, Cédric Sueur

# Impact of group management and transfer on individual sociality in Highland cattle (*Bos taurus*)

Sebastian Sosa[1], Marie Pelé[2], Élise Debergue[3,] Cédric Kuntz[3], Blandine Keller[3], Florian Robic[3], Flora Siegwalt-Baudin[3], Camille Richer[3], Amandine Ramos[3], Cédric Sueur[3]

[1] Anthropology Department, Sun-Yat Sen University, Guangzhou, China

[2] Ethobiosciences, Research and Consultancy Agency in Animal Well-Being and Behaviour, Strasbourg, France

[3] Université de Strasbourg, CNRS, IPHC UMR 7178, F-67000 Strasbourg, France

Corresponding author: cedric.sueur@iphc.cnrs.fr, 0033(0)88107453, IPHC UMR 7178, 23 rue Becquerel F-67000 Strasbourg, France

**Abstract:** The sociality of cattle facilitates the maintenance of herd cohesion and synchronisation, making these species the ideal choice for domestication as livestock for humans. However, livestock populations are not self-regulated, and farmers transfer individuals across different groups throughout their lives for reasons such as genetic mixing, reproduction and pastureland management. Individuals consequently have to adapt to different group compositions during their lives rather than choose their own herd mates, as they would do in the wild. These changes may lead to social instability and stress, entailing potentially negative effects on animal welfare. In this study, we assess how the transfer of Highland cattle (*Bos taurus*) impacts individual and group social network measures. Four groups with nine different compositions and 18 individual transfers were studied to evaluate 1) the effect of group composition on individual social centralities and 2) the effect of group composition changes on these centralities. As shown in previous works, dyadic associations are stronger between individuals with similar age and dominance rank. This study reveals that the relative stability of dyadic spatial relationships between changes in group composition or enclosure is due to the identities of transferred individuals more than the quantity of individuals that are transferred. Older cattle had higher network centralities than other individuals. The centrality of individuals was also affected by their sex and the number of familiar individuals in the group. When individuals are transferred to a group with few (one or two) or no familiar individuals, their social centralities are substantially impacted. This study reveals the necessity of understanding the social structure of a group to predict social instability following the transfer of individuals between groups. The developing of guidelines for the modification of group composition could improve livestock management and reduce stress for the animals concerned. This preprint has been peer-reviewed and recommended by Peer Community In Ecology (https://dx.doi.org/10.24072/pci.ecology.100003).

**Keywords:** livestock, social network, animal welfare, pastureland, farming, bovines

1. **Introduction**

Animal farming began in the Holocene (about 7500 years BC), when humans domesticated aurochs (*Bos primigenius*), the ancestor of *Bos Taurus*. Humans mainly chose cattle for their social nature, which facilitates the maintenance of herd cohesion and synchronisation and simplifies the locating of groups in pastureland and the coordination of movements such as transhumance (Butt et al., 2009; Von Keyserlingk et al., 2008). Social groups can regulate their own composition in the wild, with individuals migrating or groups splitting when competition for food becomes too high, for instance (Ruckstuhl and Neuhaus, 2000; Sueur et al., 2011b). This self-regulation is not possible for livestock. Farmers transfer individuals to different groups throughout their lives to facilitate genetic mixing and reproduction, or to manage pastureland activities (Bøe and Færevik, 2003; Gupta et al., 2008; Patison et al., 2010). Such changes may result in periods of social instability and stress (Estevez et al., 2007; Gutmann et al., 2015). These frequent changes in group composition modify the social organisation and stability of groups, with possible implications for animal welfare (Sueur and Pelé, 2015) and health (Costa et al., 2016).

Like their wild counterparts, domestic bovines show strong social behaviours with stable and long-term dyadic relationships when possible, i.e. when the group composition is also stable (Gutmann et al., 2015). Boyland et al. (2016) showed that cattle form strong relationships with specific partners. These preferential associations are dependent on different socio-demographic factors such as sex and age, as well as dominance, kinship or familiarity with other group members. Two individuals that are the same age or arrive in an enclosure at the same time will have a higher probability of developing a strong relationship than other individuals (Bouissou and Boissy, 2005; Bouissou et al., 2001). Many behavioural experiments have shown that cattle are able to discriminate between *familiar* and *unfamiliar individuals*, hereafter defined as individuals a bovine has spent time with, or unknown/new individuals, respectively (Hagen and Broom, 2003; Takeda et al., 2003). Adding new individuals to the group disrupts the contact between familiars and aggressive behaviour increases (Patison et al., 2010). This suggests that prioritising good and stable relationships in a group of animals enhances the wellbeing of individuals by decreasing their stress and reinforcing their social status. The use of this principle for livestock management is encouraged (Bøe and Færevik, 2003; Boyland et al., 2016; Sueur and Pelé, 2015).

In physiological terms, social stress may lead to decreased food ingestion, lower milk production and even ceased reproduction for cows (Bøe and Færevik, 2003), and can also have a strong impact on the behaviour, cognition and health of calves (Costa et al., 2016). This stress can be reduced by the presence of familiar individuals during transfer (Costa et al., 2015; Færevik et al., 2006). The impact of such transfers is also dependent on the sex of individuals: the removal of males from an enclosure leads to stronger cohesion between females, whilst the removal of females does not influence associations between males. These remain basic due to the sexual segregation observed in cattle (Ruckstuhl and Neuhaus, 2000; Wilson et al., 2015). Females are more involved in group social cohesion than males; this is probably because they are the phylopatric sex, like in some primates species (Wrangham, 1980).

It appears necessary to understand the social structure of a group to predict any social instability that could occur through the transfer of an animal. Taking this factor into consideration would make livestock management more efficient and less stressful for animals (Bøe and Færevik, 2003). This study uses social network analysis (Sueur et al., 2011a) to assess how group composition affects social centralities of Highland cattle (*Bos Taurus)* and how the transfers of these individuals impact their social relationships.

Highland cattle are originally from the Scottish Highlands in the United Kingdom. Like most domestic ungulates, this is a social species with sexual segregation (Ruckstuhl and Neuhaus, 2000). This breed is particularly suitable for eco-grazing, as it is adapted to a wide temperature range and has a non-selective diet. Many French natural reserves and national parks have imported Highland cattle in order to maintain ecosystem biodiversity (Génot, 2000; Muller et al., 1998; Wintz and Fabien, 2012). These Highland cattle populations with different group compositions can be observed in a wide study permitting a more detailed understanding of how the age ratio, sex ratio and size of group compositions affect the social centrality of cattle and how the transfer of individuals between groups impacts sociality and its dynamic in this species. We studied different compositions (nine in total) of four groups over a six-month period. We first assessed which sociodemographic factors (sex, age, dominance rank, and group size) influence the social centrality of Highland cattle, which was measured using eigenvector centrality (or popularity, i.e. how well an individual is connected to its neighbours, but also how well its neighbours are connected) and the strength of associations (or social activity, i.e. how often an individual is seen in the proximity of other specific group members) (Sueur et al., 2011a). In a second step, changes in group compositions in terms of group size, age or sex composition were examined to determine how they affected the associations and social centrality of individuals. This enabled us to measure the changes in dyadic relationships and in individual centrality according to the changes in group composition.

Following the previous results on sociality in cattle (Hagen and Broom, 2003; Reinhardt and Reinhardt, 1981; Šárová et al., 2013), we made the following hypotheses:

1. *Effects of socio-demographic factors.* Social centrality is expected to be influenced by the age, sex and dominance rank of group members and the number of familiar individuals they have in the group (Bouissou et al., 2001; Reinhardt and Reinhardt, 1981; Šárová et al., 2013, 2010; Schein and Fohrman, 1955). Older individuals were expected to have higher dominance rank and higher social centrality (Šárová et al., 2013). Familiar individuals or those of the same sex and age should also show stronger dyadic associations (Raussi et al., 2010; Ruckstuhl and Neuhaus, 2000).

2. *Effects of group composition changes*. After a transfer, fewer changes in eigenvector centrality and strength of associations were expected in older, dominant individuals, whilst the opposite was expected in younger, subordinate individuals in the new group composition. Indeed, older or dominant cattle have stronger relationships that are more easily maintained (Šárová et al., 2013, 2010). Concerning familiarity, we expected that individuals with a higher number of familiar individuals (for instance three or four) to show a lower impact on their social centrality than the individuals with no or few familiar individuals (i.e. one or two). We further predicted that resident individuals, i.e. those who experienced the arrival of a newly transferred individual in their group, would be less impacted than those being transferred (Patison et al., 2010). We suggest that the number of transferred individuals is not the only factor affecting social relationships and believe that the social role of removed or newly added individuals can have strong consequences on the social structure. We expected the removal or addition of specific individuals such as a bull or an older individual, specifically an older female, to strongly impact the social relationships of all other individuals because they no longer play their specific social role within the group (Šárová et al., 2013, 2010; Schein and Fohrman, 1955).

3. **Material & Methods**
   a. Ethical Note

This study was based on the observation of animals, and no handling or invasive experiments were involved. Our study was approved by our research institution (Institut Pluridisciplinaire Hubert Curien). It was carried out in full accordance with IPHC ethical guidelines and complied with European animal welfare legislation. Every effort was made to ensure the welfare of the animals and minimize disturbance by researchers present in the field.

   b. Observation sites and study subjects

We studied the effect of group composition and the effect of change in group composition in four groups of Highland cattle (Table 1 and Fig. 1). Group composition change is defined as changing a minority of the individuals at the study location (Robertsau, Niedersteinbach, and Sturzelbronn by either adding some new individuals or removing some individuals from the group. The four groups were located in the Grand Est region of France (see Fig. S1 for a map of the different locations). Enclosure size did not have an effect on aggression in the group or the cohesion of group members (correlation test with permutations between the enclosure size and the mean number of aggressions per day per individual: N=11, rho=-0.30, pperm=0.317; correlation test with permutations between the enclosure size and the mean number of 3m proximity per scan per individual: N=11, rho=-0.37, pperm=0.214).

Table 1: Characteristics of the four Highland cattle group sites.

| Observation site | GPS coordinates | Area (m²) | Observation time | Number of changes in group composition |
|---|---|---|---|---|
| Robertsau (Rob) | 48.611237, 7.806514 | 5 enclosure changes: 66 438;32 801; 44 028;80 501;33 637; 44 028 | Period 1: 14/04/15-28/08/15, Period 2: 22/01/16-29/04/16 | 2 |
| Niedersteinbach (Nie) | 49.029522, 7.720504 | 86 787 | Period 1: 14/04/15-28/08/15 | 1 |
| Sturzelbronn (Stu) | 49.057404, 7.580153 | 112 273 | Period 1: 14/04/15-28/08/15 | 2 |
| Rolbing (Rol) | 49.10545, 7.26120 | 71 454 | Period 2: 22/01/16-29/04/16 | None |

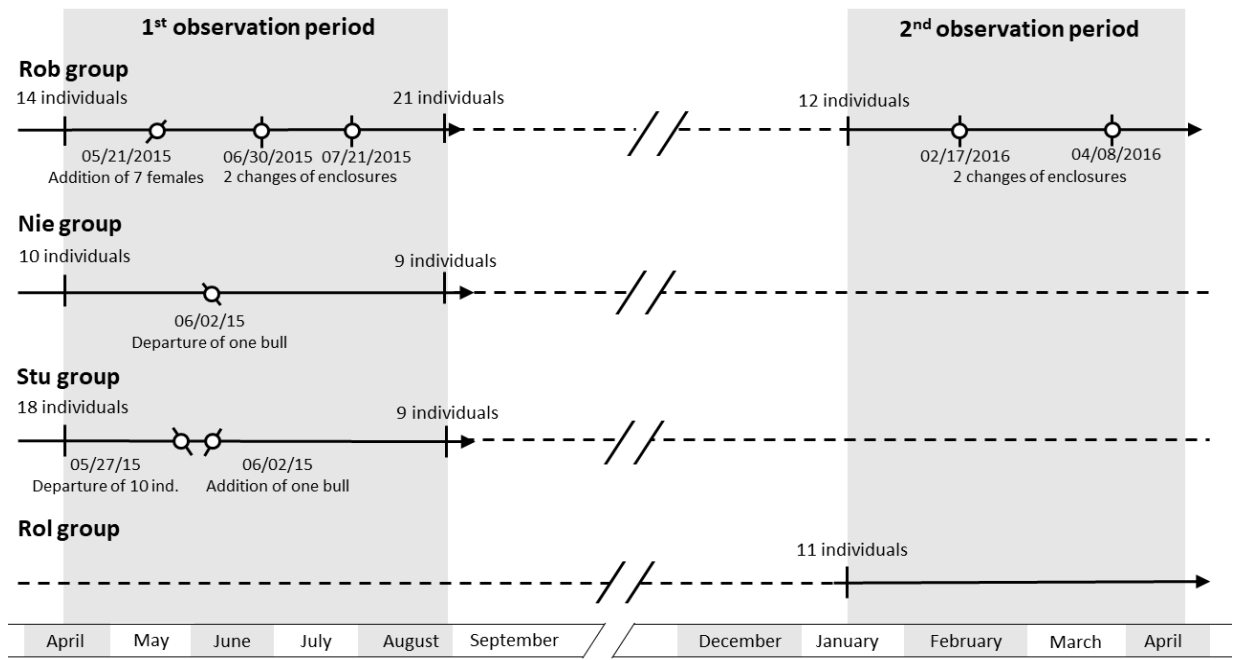

Figure 1: Chronological scheme of the composition changes in all four groups. Solid lines indicate the period of observation, whilst dashed lines indicate an absence of observation. Dots indicate changes in group composition or enclosure. Forward and backward strokes indicate the addition and departure of individuals, respectively. A vertical stroke indicates a change of enclosure.

Group composition changes were made by the farmer, either for the needs of farmland management or for breeding reasons. In particular, the non-castrated bull was transferred between the groups in order to copulate with females. Castrated bulls, which are known to be less aggressive than bulls (Bouissou, 1983; Delville et al., 1996) were also transferred into groups with juveniles to decrease the stress of the latter. Juveniles were transferred away from their mothers to facilitate new gestation. Females were generally transferred for pastureland management (Génot, 2000; Muller et al., 1998; Wintz and Fabien, 2012). The authors did not contribute to the management decision concerning the time of transfer or the choice of individuals transferred. These four groups were chosen for their group size and their contrasting group compositions (i.e. only females with juveniles, females with a bull, juveniles and bullocks; females with different versus similar ages). The groups were large enough to permit social network analysis. The group compositions were selected to study the impact of group composition on individual social centrality and how the changes of group composition affect these centralities.

Water was supplied via a water pump for the Robertsau group, whilst the three other groups had access to a river. Enclosures were all composed of similar vegetation: mainly grass (more than 90% of groundcover, surface area estimated with GIMP 2.9), wetland, some bushes and some small areas of forest/trees, as indicated in Fig. S1. Animals were supplied with hay during winter. Twice a week, hay was placed at different locations across a surface area of about seven acres to avoid resource competition. Observations were carried out over two periods: one in 2015, from April 14th to August 28th, and the second in 2016, from January 22nd to April 29th. During the two periods, composition was changed in all groups except the Rolbing group (Table 1 and Fig. 1). Each group member was identified according to physical traits such as coat colour and horn shape. These physical traits had been clearly identified for each individual prior to the study.

    **c.** Changes in group composition

Group composition changes are summarized in Fig. 1. A total of nine group compositions were observed for these four groups (Table 2) and concerned 18 individual transfers:
- Robertsau (Rob) group (3 group compositions): The group was initially composed of 14 female individuals (nine 3yo (year old) individuals and five 2yo individuals). On the 21st March 2015, seven females (two 2yo individuals and five 1yo individuals) were added to this initial group. During the second observation period, 12 individuals were removed from the group (eight 3yo individuals and four 2yo individuals) and three 2yo females were added (originating from groups other than those studied here), forming a group of 12 females (one 4yo individual, three 3yo individuals and eight 2yo individuals). We also observed four enclosure switches (change of enclosure and change of enclosure size, decided by the farmer), which had no connection with changes in group composition and were made for feeding and farmland reasons. These enclosure switches were observed on 30/06/15, 21/07/15, 17/02/16 and 08/04/16 and were taken into account in the statistical analyses.
- Niedersteinbach (Nie) group (two group compositions): the group was initially composed of ten individuals (one 7yo male, eight 5yo females and one 1yo female). The male was removed from the group on the 2nd June 2015 and was added to the Stu group.
- Sturzelbronn (Stu) group (three group compositions): the group was initially composed of 18 individuals, namely 15 females (one 13yo, two 8yo, three 7yo, two 6yo, four 2yo and three 1yo individuals) and three 1yo males. Ten individuals were removed on the 25th May 2015, namely the three 1yo males, four 2yo females and three 1yo females. An adult male (7yo) was added to the group on the 2nd June 2016.
- Rolbing (Rol) group (one group composition): the group was composed of 11 individuals, namely two castrated males (2yo), two young females (1yo) and seven young males (1yo). Its composition did not change during the study.

Table 2: Group size, number of scans and observation days, number of agonistic interactions, sex ratio and age ratio for each group composition (including changes of enclosure). For sex ratio, M= Male and F=Female. CM indicates castrated males. For age ratio, A=Adult and J=Juvenile (≤2yo).

| Group composition | Number of scans (and days) | Number of agonistic interactions | Group size | Sex ratio | Age ratio |
|---|---|---|---|---|---|
| **Niedersteinbach 1** | 429 (7) | 150 | 10 | 1 M - 9 F | 9 A - 1 J |
| **Niedersteinbach 2** | 922 (13) | 74 | 9 | 9 F | 8 A - 1 J |
| **Robertsau 1** | 207 (6) | 214 | 14 | 14 F | 9 A – 5 J |
| **Robertsau 2** | 211 (7) | 369 | 21 | 21 F | 11A – 10 J |
| **Robertsau 3** | 118 (4) | 278 | 21 | 21 F | 11A – 10 J |
| **Robertsau 4** | 221 (7) | 557 | 21 | 21 F | 11A – 10 J |
| **Robertsau 5** | 174 (4) | 104 | 12 | 12 F | 4 A - 8 J |
| **Robertsau 6** | 321 (6) | 233 | 12 | 12 F | 4 A - 8 J |
| **Robertsau 7** | 272 (4) | 99 | 12 | 12 F | 4 A - 8 J |
| **Rolbing** | 416 (9) | 74 | 11 | 7 M - 2 CM - 2 F | 2 A - 9 J |
| **Sturzelbronn 1** | 172 (4) | 83 | 18 | 3 M - 15 F | 8 A - 10 J |
| **Sturzelbronn 2** | 133 (2) | 48 | 8 | 8 F | 8 A |
| **Sturzelbronn 3** | 899 (13) | 266 | 9 | 1 M - 8 F | 1 A - 8 A |

d. Data scoring

Data were scored by two observers located two to ten meters from the animals. The two observers were always within one meter of each other. One observed and communicated the behaviours to the other, who recorded the data. This allowed behaviours to be confirmed by two observers. Cattle were already habituated to human presence and were not disturbed by the observations, which were made once a week over a six-hour period between 9am and 5pm. The groups were not observed during rainy or snowy days or during the weekends. Sampling frequency for each group composition is given in Table 2.

The group social network was defined and scored using dyadic spatial associations (Boyland et al., 2016; Sueur and Pelé, 2015). Spatial associations were defined according to the nearest neighbour (closest individual whatever the distance) and were scored every five minutes with the instantaneous sampling method (Altmann, 1974). This means that every five minutes (one scan), the value "1" was recorded in a matrix if individual A was the nearest neighbour of individual B and "0" in all other cases. We summed all scans in one matrix for each group composition, thus obtaining the absolute frequencies of nearest neighbours. Dyadic spatial association was defined as the absolute nearest neighbour frequency between each dyad of group members. The total number of scans is indicated in Table 2. The "nearest neighbour" approach is more appropriate for this kind of study (i.e. evaluating the effects of group composition on social network) than the "five meter proximity" concept (Farine and Whitehead, 2015; Franks et al., 2010). Spatial proximity matrices and nearest neighbour matrices are highly correlated (Mantel test with 1 000 permutations: $r \geq 0.78$, $p \leq 0.0001$). Given these two points, we chose the "nearest neighbour" approach to measure associations.

Observers also scored spontaneous agonistic interactions using the behavioural sampling method (Altmann, 1974) in order to assess the dominance hierarchy of each group composition. We scored supplanting, avoidance and aggression as agonistic interactions. We measured each agonistic interaction as an event, however long it lasted. We scored this interaction between individual A and individual B as "1" in a matrix of agonistic interactions. We then summed all dyadic agonistic interactions for each group composition period. Agonistic interactions, considered to be the best choice of dominance index (de Vries et al., 2006; Gammell et al., 2003), were used to calculate the Modified David's Score (MDS). David's score is based on an unweighted and a weighted sum of the individual's dyadic proportions of wins combined with an unweighted and a weighted sum of its dyadic proportions of losses (de Vries et al., 2006). Animals that usually dominate have high positive scores, and those that are usually dominated have largely negative scores. Individuals were ranked from the highest to the lowest MDS, with the individual with the highest value ranked first in the dominance hierarchy and the individual with the lowest value ranked last. SocProg 2.6 (Whitehead, 2009) was used to calculate MDS values for each group composition, and scoring began on the eighth day following transfer. We did not take the first days of observations into account in our calculation because of the instability of social and hierarchical relationships during this period. Whilst the number of aggressions were higher during these first days compared to stable periods, many agonistic behaviours were bidirectional, meaning that the hierarchy was still not established. These agonistic behaviours did not fit with the dominance ranking we observed in the stable periods.

Basing our analysis on the time intervals between group composition changes, we defined familiarity as the number of familiar individuals in the group, meaning the number of individuals a group member is with / has been with for more than three months (Sueur et al., 2017). The examination of the pedigree of each individual revealed that kinship association matrices would be difficult to obtain for each group composition due to missing data or very close genetic proximity between familiar individuals. We therefore preferred to analyse familiarity and did not assess the effect of kinship. Moreover, kinship is very difficult to study in ungulate groups, where the composition changes frequently (Færevik et al., 2006; Gutmann et al., 2015; Hagen and Broom, 2003; Patison et al., 2010; Takeda et al., 2003).

**e.** Social network analysis

Social network analysis (SNA) is an increasingly widespread tool for the study of sociality and its dynamic (Croft et al., 2008; Farine and Whitehead, 2015; Pinter-Wollman et al., 2013; Sueur et al., 2011a). Indeed, social relationships can evolve over time because of changes in the social strategies of group members, and the arrival or departure of individuals through births, deaths, migrations or transfers. Specific tools were developed in SNA to analyse these changes and their causes (Borgeaud et al., 2017, 2016; Boucherie et al., 2017; Pasquaretta et al., 2016). SNA has also been recognised as a reliable tool for animal welfare and conservation (Koene and Ipema, 2014; Snijders et al., 2017; Sueur and Pelé, 2015).

During data analysis, the matrices of spatial associations obtained per observation day were added together for each group composition. Each dyad of individuals thus obtains a spatial association weight that indicates whether or not these two individuals were frequently observed together. The spatial associations for each group composition were used to calculate the eigenvector centrality coefficient and the strength of associations of each individual (Sueur et al., 2011a). These measures were calculated using SocProg 2.6 (Whitehead, 2009).

Eigenvector centrality is a commonly used measure of individual centrality, and indicates the popularity of an individual (Kasper and Voelkl, 2009). This coefficient is defined as a measure of how well an individual is connected to its conspecifics, and also reveals the connections of the group members to which it is connected (Bonacich, 2007).

The strength of associations is the sum of each node's edge values, and indicates the social activity of an individual (Kasper and Voelkl, 2009). The individual with the strongest and most numerous associations has the highest strength value (Sueur et al., 2011a). In this study, strength indicates the number of times an individual was observed as the nearest neighbour of another individual. Indeed, in a given scan sampling, one individual might be observed several times as the nearest neighbour of the other group members (maximum = $N - 1$, where N is the group size).

These two variables are correlated but are by no means collinear (Pearson correlation test, r=0.16, p=0.03).

**f.** Statistical analyses

i. Do dyadic spatial associations depend on shared characteristics among dyads?

In a first step, we assessed how the weight of dyadic spatial associations was influenced by socio-demographic factors such as sex, age and dominance. Matrix correlations were made with a Mantel test with 1 000 permutations to check whether individuals sharing similar characteristics (similar age, dominance rank or sex) have stronger dyadic associations than individuals that do not share similar characteristics. This is called homophily, i.e. the tendency of individuals to associate and bond with similar congeners (Massen and Koski, 2014; McPherson et al., 2001). Using Socprog 2.6, we then created matrices for age differences (0: dyad individuals have the same age, 1: an age difference of approximately one year, and so on), dominance rank differences (0: dyad individuals have the same rank, 1: a difference of one dominance rank, and so on) and sex difference (0: same sex, 1: different sex). These three matrices were calculated for each group composition and correlated to the dyadic spatial association matrices for each group composition. The 'CombinePValue' package in R 3.24 was used to combine the p-value of all group compositions and obtain global statistics. The goal here was to test whether vectors of p-values are significant when combined and to confirm or negate the possible effect of a given socio-demographic factor at the population level.

ii. How does a change of group composition or enclosure affect dyadic spatial associations?

A Mantel test with 1 000 permutations in SocProg 2.6 was used to correlate the dyadic association matrices after a change (transfers or enclosure change). Only individuals that were present in the two adjacent matrices for each matrix (ex: Rob1-Rob2, Rob2-Rob3, Stu1-Stu2, etc.) were retained. The correlation coefficient was then correlated with the number of individuals transferred between two group compositions using a Spearman correlation test with permutations (library R "Coin", R 3.24).

iii. How do sociodemographic factors influence individual centralities?

GLMMs (R package 'lme4'; Bates et al., 2014) were used to test whether the eigenvector centrality and the strength of associations were affected by the following independent sociodemographic variables: the age of individuals, their sex, their dominance rank and the number of familiar individuals they were associated with in the group. The experimental units we used were the eigenvector centrality for a first GLMM and the strength of associations for a second GLMM, per individual and per group composition. Prior to GLMMs, the eigenvector centrality and the strength of associations were corrected using the group size for each composition in order to control for the mathematical effect of the number of nodes on network metrics. For the regression y=ax+b, y (the eigenvector centrality or the strength of associations) was multiplied by b. The identity of individuals was included as a random factor.

iv. How do changes in group composition affect individual centralities?

Two further GLMMs were carried out using the differences in eigenvector centrality and in strength of associations between two compositions as positive or negative values. The experimental units we used were the eigenvector centrality difference for a first GLMM and the association strength difference for a second GLMM, per individual and between two group compositions. Effect variables were the age of individuals, the number of familiar individuals in the new group, the difference in dominance rank between the two compositions (negative or positive values) and the total number of added or removed individuals. Changes of enclosures without adding or removing individuals were considered as "0" changes in the analyses. This makes it possible to compare networks where the transfer of individuals occurs to those without transfers. The identity of individuals was included as a random factor. The sex variable was not included in the model testing the differences between two group compositions because only four males (one adult and three juveniles) were transferred to another group, meaning that the sample size was too low, and the sex variable was correlated with the age of individuals in the model (male individuals were the only representatives of their age group (i.e. adult or juvenile) on transfer in all cases).

The time period was not included as random factor in our GLMMs because the variation of temperatures between the two periods (Period 1 and Period 2) was less than the difference in temperatures over a day (independent sample test with permutations: $z=4.76$, $p<0.0001$) and because the social behaviour of cattle did not change during the daytime (the changes in dyadic associations between Period 1 and Period 2 are not more numerous than the changes within each period: $r=0.6$ versus $r=0.58$). Although activity changes according to the temperature, social behaviour does not (Sueur et al., 2017). In addition, the period is not dissociated from the group composition, which has already been taken into account in our model. Taking both factors into account could lead to false interactions, influencing the statistical significance of our results (false positive or false negative, Pourhoseingholi et al., 2012).

For each GLMM, multi-model inferences were run to compare and rank candidate models according to (i) their respective Akaike Information Criterion (AIC) after correction for small sample sizes (AICc) and (ii) normalized Akaike weights (AICw) (Burnham and Anderson, 2004). Burnham and Anderson (2004) emphasized that information theoretic approaches (AIC) allow formal inference to be based on more than just one best model (lowest AIC) and lead to more robust conclusions. This means that for each combination of factors, all models were tested and ranked according to the best AIC. ΔAICc is the difference in AICc between a given model and the model with the lowest AIC. The AIC weight indicates the probability that a given model will be the best among candidate models. Models with a ΔAICc <10 were considered equally possible candidates, and their statistics were averaged. The null model (random effect: identity of individuals) was included as a possible candidate but was never among the models with lowest AICc. The results also indicate relative variable importance (RVI), which is the number of times a variable is present in the best models. Model inference and averaging were carried out with the R package 'MuMIn' (Bartoń, 2013). Node label permutations were also performed (Croft et al., 2011; Farine, 2017). Permutations are a robust and modern standard way to compare statistical models based on the original observed data to a distribution of null models based on randomised data (Farine, 2013; Farine and Whitehead, 2015). After 1000 randomisations, the statistical parameters of interest (e.g. model estimates) of the models based on observed data were compared with "null" models based on randomised data. If a substantial proportion (95%) of the statistical parameters derived from models based on observed data were lower/higher than those derived from models based on randomised data, we could conclude that the observed effects on sociality were different from those expected to arise by chance. The randomisation procedure is exactly the same for all analyses. The P-values indicated in the tables are based on these permutation procedures.

GLMM diagnostics (i.e. residual normality distribution plot and multicollinearity between dependent factors) were carried out to evaluate the validity of the final models. We checked for multicollinearity of the predictor variables by calculating the variance inflation factor (VIF, R package 'car', Fox et al., 2007). In all cases, the predictor variables had a VIF value of between 1.02 and 1.9, indicating that the predictor variables were not correlated. The significance level was set at 0.05. Statistical analyses were performed in R 3.24 (R Development Core Team, 2009). Plots of residual normality distribution can be found in the annexes (Fig. S2).

4. Results

We note that the farmer's management of cattle usually involved the transfer of young individuals. Young individuals are usually dominated by older ones in cattle (Pearson correlation test for our data: df = 176, r = -0.37, p<0.0001). Moreover, individuals arriving in a new group have fewer familiar individuals and initially have a lower dominance rank than their resident counterparts (Pearson correlation test for our data: df = 111, r = 0.41, p<0.0001), not because of their low number of familiars but because resident individuals are usually dominant over new arrivals to the group. This phenomenon is considered in the discussion.

a. Do dyadic spatial associations depend on shared characteristics among dyads?

Table 3 indicates the results of correlation tests between the dyadic association matrices and those of differences in characteristics. Fig. 2 shows four instances of Highland cattle social networks. A relatively high variability is observed according to the group composition. There is a significant correlation between matrices of dyadic associations and those of dominance rank differences. Most correlations are negative, indicating that close-ranking individuals have stronger associations than individuals with distant ranks. This is illustrated by the social networks in Fig. 2a and Fig 2b. Dyadic associations were only dependent on the sex of individuals in the Rolbing group, where individuals of the same sex had stronger associations (Fig. 2b). However, dyadic associations are mostly negatively correlated with age difference, indicating that individuals of the same age have stronger associations than cattle with greater age differences (greatest difference represented in Fig. 2c). The results for age and dominance led us to make correlations between dominance and age difference matrices. Results show that individuals of a similar age also share similar ranks; VIF analyses based at the individual level do however show that these two factors are not collinear (see Statistical Analyses in the Methods section).

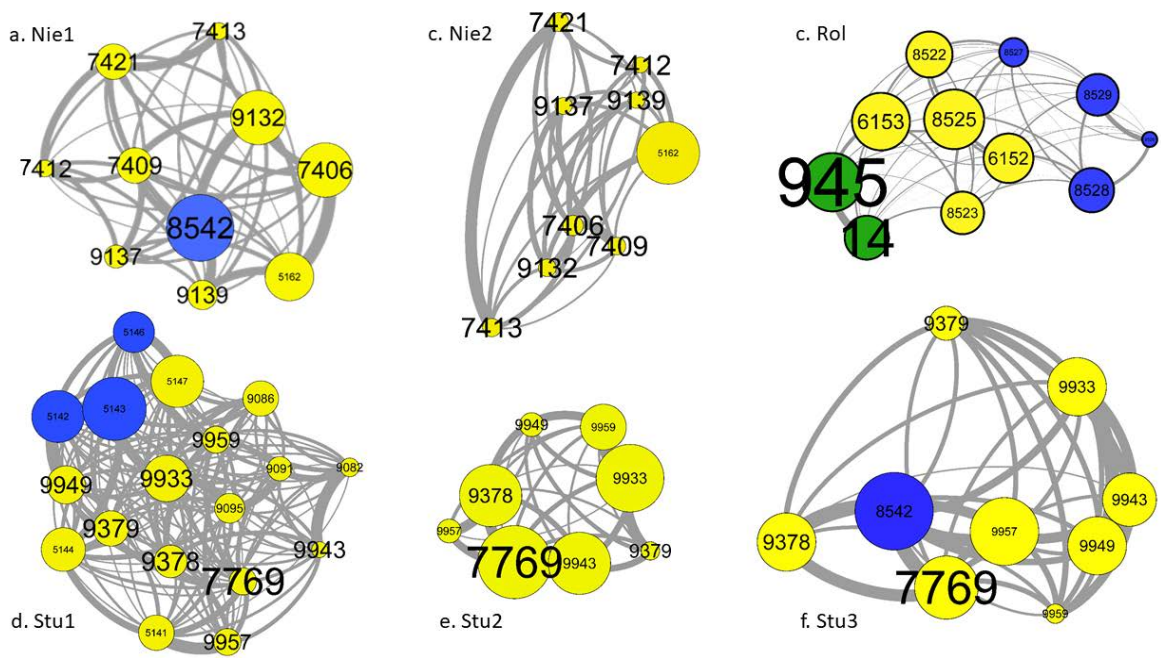

Figure 2: Examples of six group compositions: a. Nie1, b. Nie2, c. Rol, d. Stu1, e. Stu2 and f. Stu3. One node represents one individual, each identified by a number (label). The links between nodes are dyadic associations. The size of nodes depends on the strength of associations but are relative to each group composition (the strengths are not comparable between networks). The thickness of links depend on the weight of dyadic associations. The size of labels increases with the age of individuals. Yellow, blue and green node colours indicate females, males and castrated males, respectively. Individuals are positioned according to their weight of associations: two individuals located close to each other share a stronger dyadic association than distant individuals. Graphs were created using Gephi 0.91 (Bastian et al., 2009) with the 'ForceAtlas' spatialization package.

Table 3: Correlations of dyadic associations (DyaAsso) matrices with matrices of characteristic differences (age, dominance and sex). The last column also indicates the tests between matrices for age difference and dominance difference. NA = Non-Applicable. For the global value, POS indicates that most of significant correlations were positive; NEG indicates that most of significant correlations were negative.

| Group composition | DyaAsso-Dominance | DyaAsso-Sex | DyaAsso-Age | Dominance-Age |
|---|---|---|---|---|
| **Niedersteinbach 1** | p = 0.332 (r = 0.06) | p = 1 (r = -0.15) | p = 0.039 (r = 0.16) | p = 0.007 (r = 0.43) |
| **Niedersteinbach 2** | p = 0.302 (r = 0.10) | NA (just one sex) | p = 0.431 (r = -0.01) | p = 0.448 (r = 0.08) |
| **Robertsau 1** | p = 0.036 (r = -0.29) | NA (just one sex) | p = 0.004 (r = -0.15) | p = 0.002 (r = 0.52) |
| **Robertsau 2** | p<0.001 (r = -0.40) | NA (just one sex) | p<0.001 (r = -0.32) | p<0.001 (r = 0.63) |
| **Robertsau 3** | p<0.001 (r = -0.21) | NA (just one sex) | p = 0.001 (r = -0.19) | p<0.001 (r = 0.55) |
| **Robertsau 4** | p<0.001 (r = -0.40) | NA (just one sex) | p<0.001 (r = -0.32) | p<0.001 (r = 0.70) |
| **Robertsau 5** | p =0.001 (r = -0.35) | NA (just one sex) | p = 0.008 (r = -0.25) | p = 0.042 (r = 0.28) |
| **Robertsau 6** | p<0.001 (r = -0.43) | NA (just one sex) | p = 0.009 (r = -0.23) | p = 0.036 (r = 0.28) |
| **Robertsau 7** | p =0.004 (r = -0.30) | NA (just one sex) | p = 0.006 (r = -0.24) | p = 0.013 (r = 0.39) |
| **Rolbing** | p =0.015 (r = -0.30) | p = 0.006 (r = 0.31) | p<0.001 (r = -0.36) | p = 0.035 (r = 0.43) |
| **Sturzelbronn 1** | p =0.028 (r = -0.13) | p = 0.168 (r = 0.05) | p = 0.948 (r = -0.10) | p<0.001 (r = 0.63) |
| **Sturzelbronn 2** | p =0.592 (r = -0.03) | NA (just one sex) | p = 0.262 (r = 0.10) | p = 0.046 (r = 0.34) |
| **Sturzelbronn 3** | p =0.006 (r = -0.42) | p = 1 (r = -0.16) | p = 0.708 (r = -0.03) | p = 0.349 (r = 0.05) |
| **Global** | **p = 1.019e-13** **NEG (r = \|0.26\|)** | **p = 0.087** **POS(r = \|0.17\|)** | **p = 7.728e-12** **NEG(r = \|0.19\|)** | **p =2.584e-23** **POS(r = \|0.41\|)** |

b. **How does a change of group composition or enclosure affect dyadic spatial associations?**

The correlation coefficients concerning periods before and after a change ranged from -0.03 to 0.69, with an average of 0.47. This average is lower than we expected and means that 47% of relationships are stable after a change, whilst 63% change significantly. This correlation coefficient is not significantly affected by the number of transferred individuals (r=-0.49, z=-1.4, p=0.169). This result was then detailed for each group. After the removal of the male, the dyadic spatial associations of the Niedersteinbach group did not change significantly (r=0.52, p=0.0002). Dyadic spatial relationships in the Robertsau group seemed to stay stable after a change, regardless of whether if it is a change of enclosure or of group composition (0.69>r>0.52; p<0.0001). Finally, results in the Sturzelbronn group are quite different from the two previous groups with no significant stability of dyadic spatial relationships. The correlation coefficient after the removal of juveniles is -0.03 (p=0.812), and indicates the strong instability of mothers' relationships after the removal of their offspring. Similarly, the dyadic spatial relationships after the addition of the bull into the group are not significantly correlated to relationships prior to this addition (r=0.14, p=0.426), and could mean that the male has a strong impact on the relationships of females.

c. **How do sociodemographic factors influence individual centralities?**

The model selection for eigenvector centrality is indicated in Table S1. The three variables retained in the best models are dominance, familiarity and age. However, the relative importance of these variables is low (RVI(dom)=0.23; RVI(famil)=0.04; RVI(age)=0.01) and after permutations, none of these variables have a significant influence that could explain the variance of the eigenvector centrality (Table 4).

Table 4: Values of the variables retained in the best models to explain the variance of the eigenvector centrality.

|  | **Estimate** | **Std.Error** | **z value** | **Pperm left side** | **Pperm right side** |
|---|---|---|---|---|---|
| **(Intercept)** | 0.388 | 0.0157 | 24.571 | 0.00 | 1.00 |
| **Dominance** | -0.051 | 0.0204 | 2.463 | 0.199 | 0.801 |
| **Familiarity** | -0.029 | 0.026 | 1.105 | 0.298 | 0.702 |
| **Age** | 0.004 | 0.002 | 1.531 | 0.664 | 0.336 |

The model selection for the strength of associations is indicated in Table S2. The variables retained in the best models are dominance, familiarity, sex and age. Familiarity (i.e. the number of familiar individuals in the group) has a strong and significant influence on the strength of associations (RVI=0.99, Table 5, Fig. 3), i.e. the more familiars an individual has, the stronger its strength of association will be. Females also have significantly lower strengths of association than castrated males (RVI=0.89, Table 5, Fig. 4). Finally, age has a significant influence on the strength of associations (RVI=0.12, Table 5), with higher strength values in older individuals than for younger ones.

Table 5: Values of the variables retained in the best models to explain the variance of the strength of associations.

|  | Estimate | Std.Error | z value | Pperm left side | Pperm right side |
|---|---|---|---|---|---|
| (Intercept) | 0.915 | 0.138 | 6.578 | 0.65 | 0.35 |
| **Familiarity** | **0.259** | **0.065** | **3.94** | **1.00** | **0.00** |
| **SexF** | **-0.301** | **0.161** | **1.83** | **0.003** | **0.997** |
| SexM | -0.156 | 0.141 | 1.09 | 0.175 | 0.825 |
| **Age** | **0.001** | **0.005** | **0.341** | **0.98** | **0.02** |
| Dominance | -0.004 | 0.19 | 0.211 | 0.357 | 0.643 |

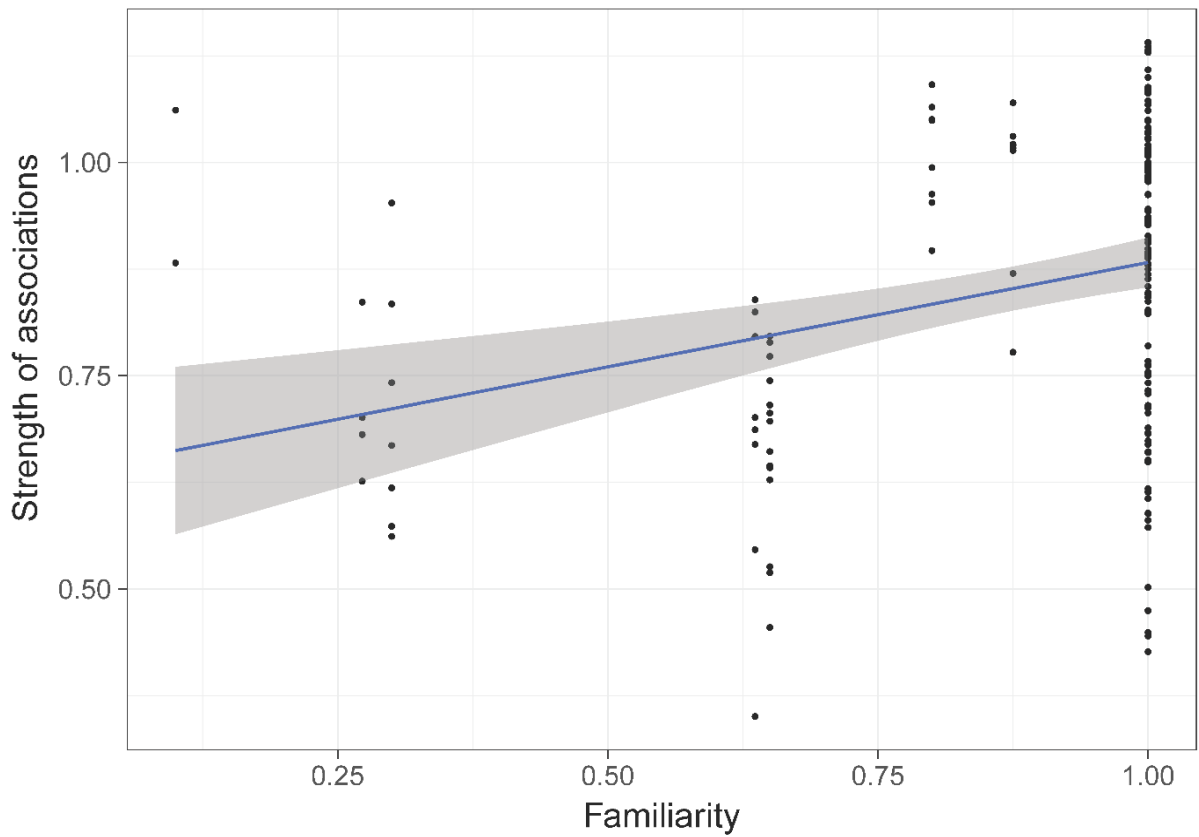

Figure 3: Strength of associations according to familiarity of individuals (i.e., proportion of familiar individuals in the group). GLMM highlighted a significant effect of familiarity on strength of associations.

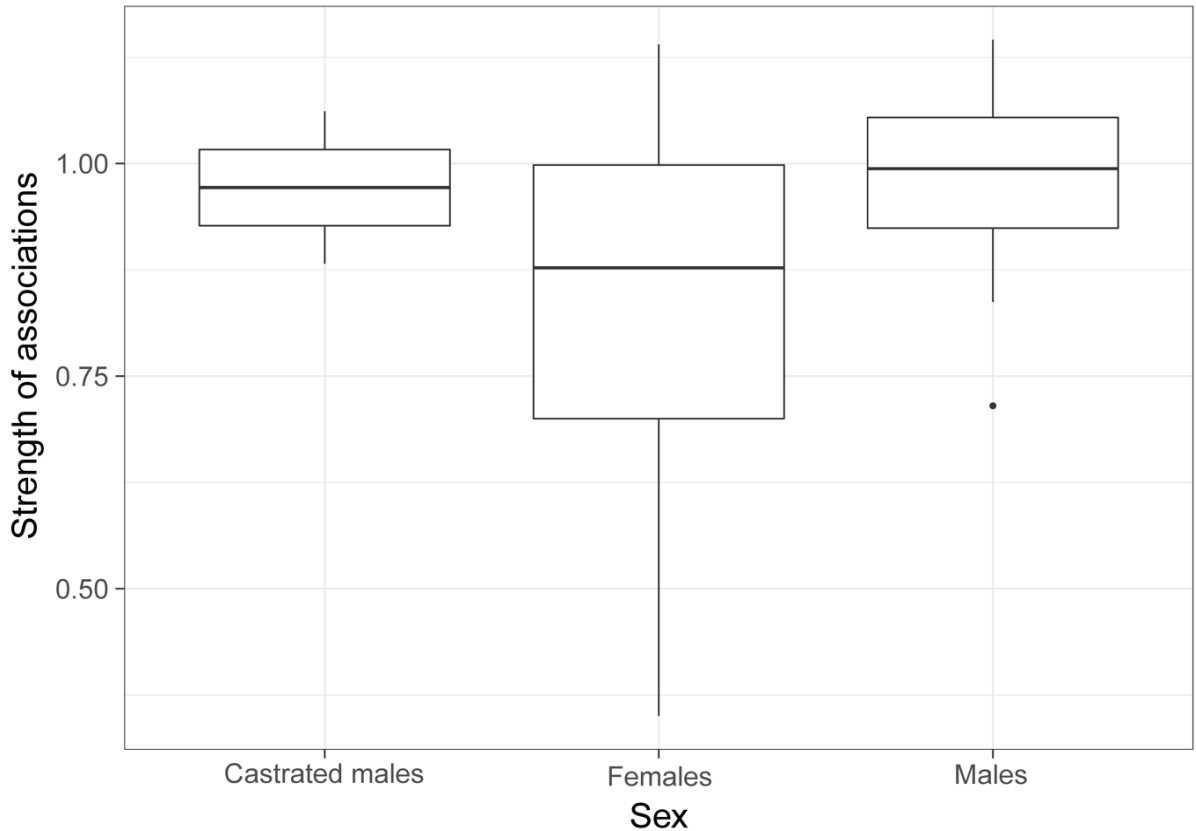

Figure 4: Strength of associations according to the sex of individuals. GLMM reveals that only the strength of associations of castrated males is different to those of females.

### d. How do changes in group composition affect individual centralities?

The model selection for the difference of eigenvector centrality after a transfer is indicated in Table S3. The three variables retained in the best models are dominance, familiarity and age. However, only age has a significant influence (RVI=0.05, Table 6), with the eigenvector centrality of older individuals increasing whilst that of younger individuals decreases (Fig. 5).

Table 6: Values of the variables retained in the best models to explain the variance of the difference of eigenvector centrality after transfer.

|  | Estimate | Std.Error | z value | Pperm left side | Pperm right side |
|---|---|---|---|---|---|
| (Intercept) | -0.010 | 0.016 | 0.629 | 0.344 | 0.656 |
| **Age** | **0.009** | **0.003** | **2.38** | **0.985** | **0.015** |
| Familiarity | -0.018 | 0.548 | 0.33 | 0.438 | 0.562 |
| Dominance | 0.017 | 0.042 | 0.392 | 0.398 | 0.602 |

The model selection for the difference of strength of associations after a transfer is indicated in Table S5. The variables retained in the best models are dominance, familiarity in the new group, age, and the number of transferred individuals. However, only the number of familiar individuals in the new group had a significant influence on the difference of strength of associations (RVI=1, Table 7), with individuals that had greater numbers of familiar individuals showing stronger strengths of association (Fig. 6).

Table 7: Values of the variables retained in the best models to explain the variance of the difference of strength of associations after transfer.

|  | Estimate | Std.Error | Adjusted SE | z value | Pperm left side | Pperm right side |
|---|---|---|---|---|---|---|
| (Intercept) | -0.730 | 0.145 | 0.147 | 4.975 | 0.00 | 1.00 |
| Dominance | 0.120 | 0.143 | 0.144 | 0.831 | 0.90 | 0.10 |
| **Familiarity** | **0.816** | **0.151** | **0.153** | **5.323** | **1.00** | **0.00** |
| Age | 0.001 | 0.005 | 0.005 | 0.213 | 0.812 | 0.188 |
| N | -0.0003 | 0.002 | 0.002 | 0.159 | 0.112 | 0.888 |

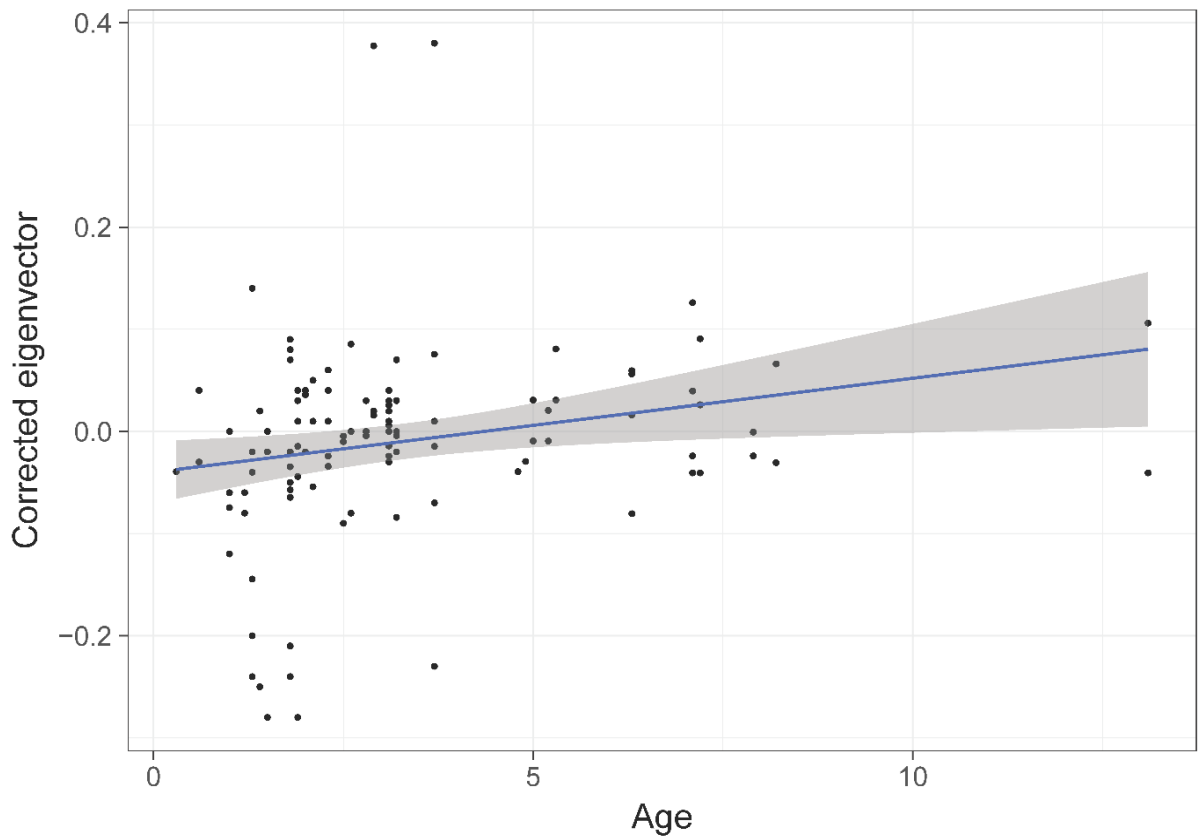

Figure 5: Difference of eigenvectors after a transfer, according to the age of individuals. GLMM highlights a significant effect of age of individuals on the change in strength of associations after a transfer.

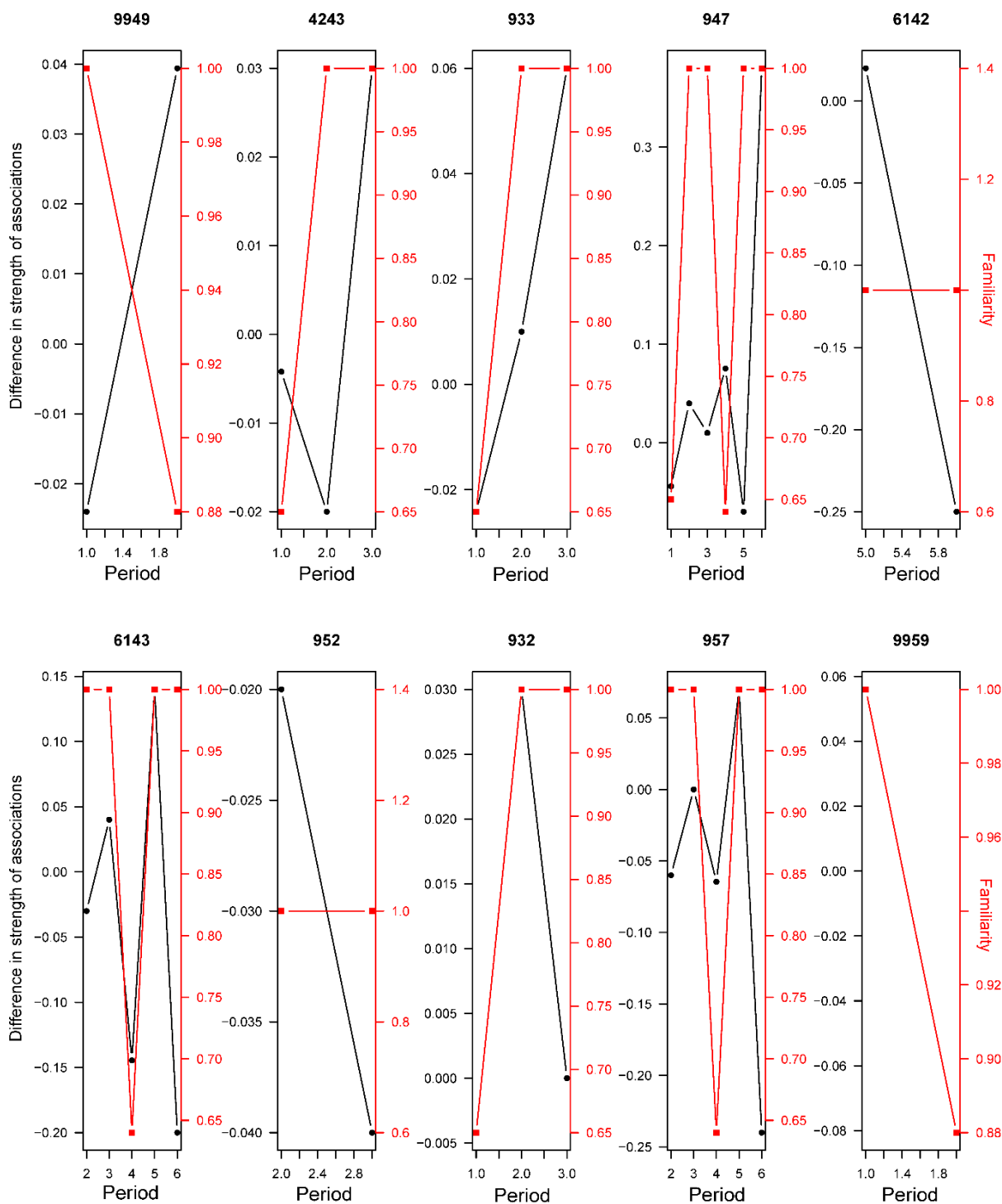

Figure 6: Difference in strength of associations (black line) and familiarity (red line, as proportion of familiar individuals in the group) between different periods of transfer (Periods exclude any transfer activity). Examples for ten randomly chosen individuals.

## 5. Discussion

This study shows how individual and dyadic social network metrics are shaped by sociodemographic factors and composition changes in several groups of Highland cattle. Analyses of dyadic associations and individual centralities highlighted correlations between spatial proximity, age and dominance, an influence of familiarity, age and sex on individual centralities, and finally an impact of transfers that mainly varied according to the number of individuals with which the transferred animal was already familiar.

### a. Do dyadic spatial associations depend on shared characteristics among dyads?

Matrix correlation tests revealed that individuals of similar age and dominance rank develop stronger associations and are located closer to each other than individuals of different age and sex. However, the tests also showed a correlation between age and dominance rank similarities. For instance, individuals 951, 949 and 947 in the Robertsau 6 group composition (Fig 2a) are approximately the same age, are the top-ranking individuals and form a triad with strong associations. This configuration has also been reported in female mouflons (*Ovis gmelini*) where the most dominant females form triadic relationships (Guilhem et al., 2002; Le Pendu et al., 2000) and is reminiscent of "triadic closure", a mechanism that may facilitate the development of cooperation for social alliances or access to food. However, it is not clear whether triadic closure is a by-product of socio-demographic characteristics (i.e. individuals that share the same characteristics also share the same needs), or if it is a social strategy leading to better cooperation between multiple partners (Banks and Carley, 1996; Righi and Takacs, 2014). Other examples also show this homophily according to age and dominance (Lusseau and Newman, 2004; Massen and Koski, 2014). Many authors have confirmed homophily (tendency of individuals to associate and bond with similar others) in ungulate species (Gerard and Richard-Hansen, 1992; Guilhem et al., 2000; Kimura, 1998; Roberts and Browning, 1998; Winfield et al., 1981), and underline that animals with the same socio-demographic characteristics may also share the same social or physiological/nutritional needs. Indeed, younger individuals show strong associations, as observed in the Rolbing and Sturzelbronn 1 group compositions (Fig. 2b and Fig. 2c, respectively). This homophily seems to help young individuals to learn how to live in groups and acquire sociality without risk of injury, particularly when in contact with adults (Shimada and Sueur, 2014). The same reasoning about reducing risk of injury could be applied for homophily between individuals that have the same dominance rank. Risk of injury prevents subordinate individuals from having strong associations with dominant individuals (as described in ungulates by Syme et al. 1975 and in primates by Balasubramaniam et al., 2012; Borgeaud et al., 2016; Sosa, 2016). This dominance-related homophily may also result from competition between individuals seeking to associate with top-ranking individuals on order to obtain tolerance or access to resources. However, as high-ranking individuals are already associated among themselves, low-ranking individuals might not gain access to them (Borgeaud et al., 2016). The results we obtained were not observed in all group compositions, and this could be explained by intra-group age variance. The difference in dominance and the strength of homophily increase with differences in age. This was seen in the Niedersteinbach group, where the maximum age difference between individuals was two years (individuals aged 7yo and 5yo, with the exception of one juvenile). Unlike the other compositions, no age-related homophily was observed in this group.

Individuals of the same age also have more similar dominance ranks than individuals of different ages. Age affects dominance through the association of individuals, meaning that individuals of the same age are likely to develop the same dominance rank because of their strong and close associations. Social status such as dominance increases with age through different processes such as increases in body weight, experience and knowledge or social power (Crockford, 2016; McComb et al., 2011; Šárová et al., 2013; Sosa, 2016; Tokuyama and Furuichi, 2017). In the Niedersteinbach 1 group composition, the male, which was also the oldest and highest-ranking individual, played an important role in the correlation with dyadic associations. The correlation was no longer significant when this individual left the group (Niedersteinbach 2). This is either simply because it had been removed from the statistics, or because the group's social structure had been perturbed. When this male arrived in the Sturzelbronn 3 group composition, it was no longer the oldest in the group but it became the highest ranking individual, making the correlation with dominance and associations significant.

Whilst age and dominance have a strong impact on dyadic relationships, we found that age was the only variable affecting strength of associations. Older individuals obtain stronger strengths of associations, but dominant individuals do not. There does not appear to be any competition for the central positions in the groups we studied. Dominant individuals are usually expected to develop strong associations because they occupy central positions in the group for better protection against predators or increased access to other resources. This affords higher centrality to these dominant individuals than to others. Other resources are used in this system, such as small clumps of trees that protect from the sun and high temperatures. These spots are appreciated by animals for thermoregulation, and dominant individuals have been seen to occupy them and prevent others from entering them (Laforge et al., 2016; Lopes et al., 2016; McCann et al., 2016). Whilst some such areas were present in our study groups, no correlation of this type was observed between strength of associations and dominance.

**b. How do sociodemographic factors influence individual centralities?**

Centrality is also linked to age, with the oldest individuals having the highest strength of associations. With age, individuals become more and more selective (Almeling et al., 2016) in their social relationships. Young individuals interact unselectively with many partners in order to learn social rules (Shimada and Sueur, 2014). With time, they develop more stable relationships and become more and more central (Sosa, 2016). In our study, this effect was amplified because young individuals, juveniles or young adults were also those the farmers chose to transfer. They therefore had to develop new relationships each time they were transferred, accentuating the link between age and centrality. Juveniles usually have strong relationships with their mothers, yet few juveniles were still in the presence of their mother in our study. They were not easily accepted on their transferal and remained on the periphery of the new group, forming strong dyadic associations among themselves as already shown in previous studies (Bøe and Færevik, 2003; Raussi et al., 2010). This result for age is emphasised by that obtained for familiarity. Indeed, in our study, familiarity was linked to age as older individuals stayed in their enclosure whilst younger ones were transferred. Individuals with a greater number of familiar individuals in the group showed higher centralities. In bovines, group members form subsets of familiar individuals, accentuating dyadic relationships and increasing centralities (Gutmann et al., 2015; Sato et al., 1993). In sheep (*Ovies aries*), familiar individuals are attracted to each other, whilst non-familiar individuals are not (Winfield et al., 1981). In our study, resident cattle rebuffed new individuals and were more aggressive towards them around coveted spots (personal observations). The same result has been found in barnacle geese (*Branta leucopsis*) (Kurvers et al., 2013) and in mallards (*Anas platyrhynchos*) (Bousquet et al., 2017).

Sex also affected the strength of associations in our study, with castrated males showing stronger strengths of association than females. We did not observe any difference between males and females, and this is mainly due to the social organisation of bovines. Bovines show sexual segregation, and females usually develop stronger and more stable dyadic associations than males, resulting in a higher centrality for females (Bouissou et al., 2001; Reinhardt and Reinhardt, 1981; Villaret and Bon, 1998). However, this sex-centrality link in our study is influenced by the fact that male juveniles remain closer to their mother and other young individuals. The stronger centralities of the two castrated males in our study are mainly explained by the group composition. These two individuals were the two only adults in a group of juveniles, which seek group cohesion more than adults. The sex variable was therefore not dissociated from with age in the Rolbing group, which probably explains this result in our study.

Eigenvector centrality was not affected by any of the factors we studied. This is probably because eigenvector centrality takes not only direct connections (i.e how an individual is connected) into account, but also indirect connections, i.e. how its neighbours are connected to other individuals (Bonacich, 2007). Our studied groups were quite cohesive with a low sample size, which may have led to a low variance of eigenvector centralities between group members and an absence of correlations with socio-demographic factors.

Together, these results allow us to identify which factors affect the social relationships and thus the centralities of group members; the combination of these factors as a management tool could reinforce group cohesion by giving a key sociality role to one specific group member or decreasing aggressiveness during group transfers.

### c. How does a change of group composition or of enclosure affect dyadic spatial associations?

Our results showed that social relationships are more affected by the identities of transferred individuals than by the number of individuals transferred. Indeed, the addition or the removal of young or adult individuals that were not related to other group members does not seem to significantly affect the social relationships of resident individuals, except for the addition of the male in Sturzelbronn. However, the removal of offspring seems to strongly destabilize the relationships of the mothers.

### d. How do changes in group composition affect individual centralities?

The difference in the eigenvector centralities between two transfers is explained by age alone. Results show that the centrality of young individuals tends to decrease during transfer, whilst older individuals obtain higher centrality. During transfer, most young individuals leave their original group for a new group without their mother. These individuals are then isolated and placed at the periphery of the group until they form new and stable relationships (Færevik et al., 2006). Conversely, adults benefit from the transfer of young individuals as they are residents, and newly transferred individuals seek cohesion to alleviate their stress. Indeed, stress increases social cohesion and proximity with partners (Dufour et al., 2011; Hansen et al., 2009; Heathcote et al., 2017). Moreover, the eigenvector centrality coefficient takes into account not only the connections of a group member, but also how these connections are connected to other individuals in the group (Bonacich, 2007). If the relationships of an individual change but those of its connected individuals do not, then little change will be seen in eigenvector centrality, whilst the strength of association will increase or decrease. In this respect, the eigenvector centrality coefficient is more stable than coefficients that are solely focused on the individual, such as strength of associations or degree (Levé et al., 2016).

Strength of associations was only affected by the number of familiar individuals in the new group. Individuals with a stable number of familiar individuals in the new group composition showed frequent interaction with them, whilst the individuals that had been separated from familiar individuals interacted less with other group members and needed time to develop strong and stable associations. Færevik et al. (2006) showed that the presence of familiar individuals during transfer is indeed less stressful. Familiar individuals have a stronger strength of associations due to increased group cohesion (Dufour et al., 2011; Hansen et al., 2009). Finally, and surprisingly, the number of transferred individuals did not lead to a change in strength of associations. Mathematically, as there are more partners to associate with when the number of individuals increases in a group, there is less possibility and less time for each partner to associate. We should therefore observe a global decrease in the strength of associations per individual. Kondo et al. (1989) showed that an increase in group size may lead to decreased space availability and therefore result in a higher occurrence of agonistic behaviours. The fact that we did not observe such an effect in our study, at least after removing the first eight days after a transfer, could be explained by the large size of the enclosures. Indeed, aggressive interactions are at their highest when the groups are first mixed. In most cattle groups, aggression is rarely seen once the dominance rank is established, as groups operate more through affiliative than agonistic behaviours (Schein and Fohrman, 1955). Newly transferred individuals in this study have usually all been removed from the same group, which may lead these individuals to stay together (resident vs. transferred) and thus exclude any change in their relationships. However, this hypothesis remains to be tested as even if they stay amongst themselves, the stress entailed by the change should lead to a greater cohesion of individuals, and this was not observed in our study.

### e. Implication for animal welfare

Our results show that a group is structured according to age, dominance and familiarity. Favouring specific age differences between individuals and subsets of familiars may be a tool to control cohesion and stability and decrease aggression in a group. The individual centralities of cattle decrease during transfers and changes in group composition. This occurs mainly in young individuals and is due to the loss of familiar individuals. During stressful events, animals seem to prefer interacting with familiar individuals and avoid interacting with unfamiliar group members (Winfield et al., 1981). When transferring individuals, it is therefore preferable to select a certain number of familiar individuals to transfer as a group in order to decrease stress. Although it is true that animals should adapt to their new environment after a certain time (Estevez et al., 2007) an optimal group composition will permit a more rapid integration of new individuals. This is particularly important in view of the fact that stress can impact the behaviour, cognition, reproductive performance and health of individuals (Costa et al., 2016; Gaillard et al., 2014; Proudfoot and Habing, 2015). It would also be preferable to transfer juvenile individuals aged around 3yo with an adult, and avoid transferring juveniles that are less than one year old. This would be the best way to decrease the stress of juveniles to a minimum during transfer. On the other hand, forming stable pairs of individuals before and during transfers may increase food intake and weight gain, particularly in calves (Costa et al., 2015). This study has highlighted some interesting results for the improvement of livestock welfare, but other factors could be studied to further enhance animal wellbeing during changes in group composition, notably the personality of the individuals chosen for transfer (Pruitt and Keiser, 2014; Wolf and Weissing, 2012).

**Acknowledgements**

We would like to thank Cécile Bayeur and Jean-Claude Bieber from the "Parc Naturel des Vosges du Nord", France and Frédéric Lonchampt from Strasbourg, France for granting us access to the Highland cattle. Our thanks to Arthur Letzelter and Fabien Kauffer for their help obtaining data for the socio-demographic factors of animals, and to Louise Frey and Audrey Claus for their help in data scoring. We thank Joanna Lignot (Munro Language Services) for English editing and Alecia Carter for her helpful comments. Sebastian Sosa is funded by the National Natural Science Foundation of China (31470456).

# References

Almeling, L., Hammerschmidt, K., Sennhenn-Reulen, H., Freund, A.M., Fischer, J., 2016. Motivational shifts in aging monkeys and the origins of social selectivity. Curr. Biol. 26, 1744–1749.
Altmann, J., 1974. Observational Study of Behavior: Sampling Methods. Behaviour 49, 227–267.
Balasubramaniam, K.N., Dittmar, K., Berman, C.M., Butovskaya, M., Cooper, M.A., Majolo, B., Ogawa, H., Schino, G., Thierry, B., De Waal, F.B.M., 2012. Hierarchical steepness, counter-aggression, and macaque social style scale. Am. J. Primatol. 74, 915–925. https://doi.org/10.1002/ajp.22044
Banks, D.L., Carley, K.M., 1996. Models for network evolution. J. Math. Sociol. 21, 173–196.
Bartoń, K., 2013. MuMIn: multi-model inference. R Package Version 1.
Bastian, M., Heymann, S., Jacomy, M., 2009. Gephi: An Open Source Software for Exploring and Manipulating Networks.
Bates, D., Maechler, M., Bolker, B., Walker, S., 2014. lme4: Linear mixed-effects models using Eigen and S4. R Package Version 1, 1–23.
Bøe, K.E., Færevik, G., 2003. Grouping and social preferences in calves, heifers and cows. Appl. Anim. Behav. Sci. 80, 175–190.
Bonacich, P., 2007. Some unique properties of eigenvector centrality. Soc. Netw. 29, 555–564. https://doi.org/10.1016/j.socnet.2007.04.002
Borgeaud, C., Sosa, S., Bshary, R., Sueur, C., van de Waal, E., 2016. Intergroup Variation of Social Relationships in Wild Vervet Monkeys: A Dynamic Network Approach. Front. Psychol. 7, 915. https://doi.org/10.3389/fpsyg.2016.00915
Borgeaud, C., Sosa, S., Sueur, C., Bshary, R., 2017. The influence of demographic variation on social network stability in wild vervet monkeys. Anim. Behav. 134, 155–165. https://doi.org/10.1016/j.anbehav.2017.09.028
Boucherie, P.H., Sosa, S., Pasquaretta, C., Dufour, V., 2017. A longitudinal network analysis of social dynamics in rooks corvus frugilegus: repeated group modifications do not affect social network in captive rooks. Curr. Zool. 63, 379–388. https://doi.org/10.1093/cz/zow083
Bouissou, M., Boissy, A., 2005. Le comportement social des bovins et ses conséquences en élevage. INRA Prod. Anim. 18, 87–99.
Bouissou, M.F., 1983. Androgens, aggressive behaviour and social relationships in higher mammals. Horm. Res. 18, 43–61.
Bouissou, M.-F., Boissy, A., Le Neindre, P., Veissier, I., 2001. The social behaviour of cattle. Soc. Behav. Farm Anim. 113–145.
Bousquet, C.A.H., Ahr, N., Sueur, C., Petit, O., 2017. Determinants of leadership in groups of female mallards. https://doi.org/10.1163/1568539X-00003431
Boyland, N.K., Mlynski, D.T., James, R., Brent, L.J.N., Croft, D.P., 2016. The social network structure of a dynamic group of dairy cows: From individual to group level patterns. Appl. Anim. Behav. Sci. 174, 1–10. https://doi.org/10.1016/j.applanim.2015.11.016
Burnham, K.P., Anderson, D.R., 2004. Multimodel inference understanding AIC and BIC in model selection. Sociol. Methods Res. 33, 261–304.
Butt, B., Shortridge, A., WinklerPrins, A.M., 2009. Pastoral herd management, drought coping strategies, and cattle mobility in southern Kenya. Ann. Assoc. Am. Geogr. 99, 309–334.
Costa, J., Meagher, R., von Keyserlingk, M., Weary, D., 2015. Early pair housing increases solid feed intake and weight gains in dairy calves. J. Dairy Sci. 98, 6381–6386.
Costa, J., von Keyserlingk, M., Weary, D., 2016. Invited review: Effects of group housing of dairy calves on behavior, cognition, performance, and health. J. Dairy Sci. 99, 2453–2467.
Crockford, C., 2016. Aging: Lessons for Elderly People from Monkeys. Curr. Biol. 26, R532–R534.
Croft, D.P., James, R., Krause, J., 2008. Exploring Animal Social Networks. Princeton University Press.
Croft, D.P., Madden, J.R., Franks, D.W., James, R., 2011. Hypothesis testing in animal social networks. Trends Ecol. Evol. 26, 502–507. https://doi.org/10.1016/j.tree.2011.05.012

de Vries, H., Stevens, J.M.G., Vervaecke, H., 2006. Measuring and testing the steepness of dominance hierarchies. Anim. Behav. 71, 585–592. https://doi.org/10.1016/j.anbehav.2005.05.015

Delville, Y., Mansour, K.M., Ferris, C.F., 1996. Testosterone facilitates aggression by modulating vasopressin receptors in the hypothalamus. Physiol. Behav. 60, 25–29.

Dufour, V., Sueur, C., Whiten, A., Buchanan-Smith, H. m., 2011. The impact of moving to a novel environment on social networks, activity and wellbeing in two new world primates. Am. J. Primatol. 73, 802–811. https://doi.org/10.1002/ajp.20943

Estevez, I., Andersen, I.-L., Nævdal, E., 2007. Group size, density and social dynamics in farm animals. Appl. Anim. Behav. Sci. 103, 185–204.

Færevik, G., Jensen, M.B., Bøe, K.E., 2006. Dairy calves social preferences and the significance of a companion animal during separation from the group. Appl. Anim. Behav. Sci. 99, 205–221.

Farine, D.R., 2013. Animal social network inference and permutations for ecologists in R using asnipe. Methods Ecol. Evol. 4, 1187–1194. https://doi.org/10.1111/2041-210X.12121

Farine, D.R., n.d. A guide to null models for animal social network analysis. Methods Ecol. Evol. n/a-n/a. https://doi.org/10.1111/2041-210X.12772

Farine, D.R., Whitehead, H., 2015. Constructing, conducting and interpreting animal social network analysis. J. Anim. Ecol. 84, 1144–1163.

Fox, J., Friendly, G.G., Graves, S., Heiberger, R., Monette, G., Nilsson, H., Ripley, B., Weisberg, S., Fox, M.J., Suggests, M., 2007. The car package.

Franks, D.W., Ruxton, G.D., James, R., 2010. Sampling animal association networks with the gambit of the group. Behav. Ecol. Sociobiol. 64, 493–503. https://doi.org/10.1007/s00265-009-0865-8

Gaillard, C., Meagher, R.K., von Keyserlingk, M.A., Weary, D.M., 2014. Social housing improves dairy calves' performance in two cognitive tests. PloS One 9, e90205.

Gammell, M.P., Vries, H. de, Jennings, D.J., Carlin, C.M., Hayden, T.J., 2003. David's score: a more appropriate dominance ranking method than Clutton-Brock et al.'s index [WWW Document]. Anim. Behav. URL http://dspace.library.uu.nl/handle/1874/17476 (accessed 2.18.17).

Génot, J.-C., 2000. Conservation de la nature: gérer les espèces ou les habitats? Le cas du parc naturel régional des Vosges du Nord, réserve de la biosphère. Courr. Environ. INRA 5–18.

Gerard, J.-F., Richard-Hansen, C., 1992. Social affinities as the basis of the social organization of a Pyrenean chamois (Rupicapra pyrenaica) population in an open mountain range. Behav. Processes 28, 111–122.

Guilhem, C., Bideau, E., Gerard, J., Maublanc, M., 2000. Agonistic and proximity patterns in enclosed mouflon (Ovis gmelini) ewes in relation to age, reproductive status and kinship. Behav. Processes 50, 101–112.

Guilhem, C., Gerard, J., Bideau, E., 2002. Rank acquisition through birth order in mouflon sheep (Ovis gmelini) ewes. Ethology 108, 63–73.

Gupta, S., Earley, B., Nolan, M., Formentin, E., Crowe, M.A., 2008. Effect of repeated regrouping and relocation on behaviour of steers. Appl. Anim. Behav. Sci. 110, 229–243.

Gutmann, A.K., Špinka, M., Winckler, C., 2015. Long-term familiarity creates preferred social partners in dairy cows. Appl. Anim. Behav. Sci. 169, 1–8.

Hagen, K., Broom, D.M., 2003. Cattle discriminate between individual familiar herd members in a learning experiment. Appl. Anim. Behav. Sci. 82, 13–28.

Hansen, H., McDonald, D.B., Groves, P., Maier, J.A.K., Ben-David, M., 2009. Social Networks and the Formation and Maintenance of River Otter Groups. Ethology 115, 384–396. https://doi.org/10.1111/j.1439-0310.2009.01624.x

Heathcote, R.J.P., Darden, S.K., Franks, D.W., Ramnarine, I.W., Croft, D.P., 2017. Fear of predation drives stable and differentiated social relationships in guppies. Sci. Rep. 7, 41679. https://doi.org/10.1038/srep41679

Kasper, C., Voelkl, B., 2009. A social network analysis of primate groups. Primates 50, 343–356. https://doi.org/10.1007/s10329-009-0153-2

Kimura, R., 1998. Mutual grooming and preferred associate relationships in a band of free-ranging horses. Appl. Anim. Behav. Sci. 59, 265–276.

Koene, P., Ipema, B., 2014. Social Networks and Welfare in Future Animal Management. Animals 4, 93–118. https://doi.org/10.3390/ani4010093

Kondo, S., Sekine, J., Okubo, M., Asahida, Y., 1989. The effect of group size and space allowance on the agonistic and spacing behavior of cattle. Appl. Anim. Behav. Sci. 24, 127–135.

Kurvers, R.H., Adamczyk, V.M., Kraus, R.H., Hoffman, J.I., van Wieren, S.E., van der Jeugd, H.P., Amos, W., Prins, H.H., Jonker, R.M., 2013. Contrasting context dependence of familiarity and kinship in animal social networks. Anim. Behav. 86, 993–1001.

Laforge, M.P., Michel, N.L., Wheeler, A.L., Brook, R.K., 2016. Habitat selection by female moose in the Canadian prairie ecozone. J. Wildl. Manag. 80, 1059–1068.

Le Pendu, Y., Guilhem, C., Briedermann, L., Maublanc, M.-L., Gerard, J.-F., 2000. Interactions and associations between age and sex classes in mouflon sheep (Ovis gmelini) during winter. Behav. Processes 52, 97–107.

Levé, M., Sueur, C., Petit, O., Matsuzawa, T., Hirata, S., 2016. Social grooming network in captive chimpanzees: does the wild or captive origin of group members affect sociality? Primates 57, 73–82.

Lopes, L.B., Eckstein, C., Pina, D.S., Carnevalli, R.A., 2016. The influence of trees on the thermal environment and behaviour of grazing heifers in Brazilian Midwest. Trop. Anim. Health Prod. 48, 755–761.

Lusseau, D., Newman, M.E.J., 2004. Identifying the role that animals play in their social networks. Proc. R. Soc. B Biol. Sci. 271, S477–S481. https://doi.org/10.1098/rsbl.2004.0225

Massen, J.J., Koski, S.E., 2014. Chimps of a feather sit together: chimpanzee friendships are based on homophily in personality. Evol. Hum. Behav. 35, 1–8.

McCann, N.P., Moen, R.A., Windels, S.K., Harris, T.R., 2016. Bed sites as thermal refuges for a cold-adapted ungulate in summer. Wildl. Biol. 22, 228–237.

McComb, K., Shannon, G., Durant, S.M., Sayialel, K., Slotow, R., Poole, J., Moss, C., 2011. Leadership in Elephants: The Adaptive Value of Age. Proc. R. Soc. B Biol. Sci. https://doi.org/10.1098/rspb.2011.0168

McPherson, M., Smith-Lovin, L., Cook, J.M., 2001. Birds of a feather: Homophily in social networks. Annu. Rev. Sociol. 27, 415–444.

Muller, S., Dutoit, T., Alard, D., Grevilliot, F., 1998. Restoration and Rehabilitation of Species-Rich Grassland Ecosystems in France: a Review. Restor. Ecol. 6, 94–101.

Pasquaretta, C., Klenschi, E., Pansanel, J., Battesti, M., Mery, F., Sueur, C., 2016. Understanding Dynamics of Information Transmission in Drosophila melanogaster Using a Statistical Modeling Framework for Longitudinal Network Data (the RSiena Package). Front. Psychol. 7. https://doi.org/10.3389/fpsyg.2016.00539

Patison, K.P., Swain, D.L., Bishop-Hurley, G.J., Robins, G., Pattison, P., Reid, D.J., 2010. Changes in temporal and spatial associations between pairs of cattle during the process of familiarisation. Appl. Anim. Behav. Sci. 128, 10–17.

Pinter-Wollman, N., Hobson, E.A., Smith, J.E., Edelman, A.J., Shizuka, D., Silva, S. de, Waters, J.S., Prager, S.D., Sasaki, T., Wittemyer, G., Fewell, J., McDonald, D.B., 2013. The dynamics of animal social networks: analytical, conceptual, and theoretical advances. Behav. Ecol. art047. https://doi.org/10.1093/beheco/art047

Pourhoseingholi, M.A., Baghestani, A.R., Vahedi, M., 2012. How to control confounding effects by statistical analysis. Gastroenterol. Hepatol. Bed Bench 5, 79–83.

Proudfoot, K., Habing, G., 2015. Social stress as a cause of diseases in farm animals: current knowledge and future directions. Vet. J. 206, 15–21.

Pruitt, J.N., Keiser, C.N., 2014. The personality types of key catalytic individuals shape colonies' collective behaviour and success. Anim. Behav. 93, 87–95.

R Development Core Team, 2009. R: A language and environment for statistical computing. R Foundation for Statistical Computing, Vienna, Austria.

Raussi, S., Niskanen, S., Siivonen, J., Hänninen, L., Hepola, H., Jauhiainen, L., Veissier, I., 2010. The formation of preferential relationships at early age in cattle. Behav. Processes 84, 726–731.

Reinhardt, V., Reinhardt, A., 1981. Cohesive Relationships in a Cattle Herd (Bos Indicus). Behaviour 77, 121–150. https://doi.org/10.1163/156853981X00194

Righi, S., Takacs, K., 2014. Triadic balance and closure as drivers of the evolution of cooperation. Presented at the Social Simulation Conference.

Roberts, J.M., Browning, B.A., 1998. Proximity and threats in highland ponies. Soc. Netw. 20, 227–238.

Ruckstuhl, K.E., Neuhaus, P., 2000. Sexual Segregation in Ungulates: A New Approach. Behaviour 137, 361–377.

Šárová, R., Špinka, M., Panamá, J.L.A., Šimeček, P., 2010. Graded leadership by dominant animals in a herd of female beef cattle on pasture. Anim. Behav. 79, 1037–1045. https://doi.org/10.1016/j.anbehav.2010.01.019

Šárová, R., Špinka, M., Stěhulová, I., Ceacero, F., Šimečková, M., Kotrba, R., 2013. Pay respect to the elders: age, more than body mass, determines dominance in female beef cattle. Anim. Behav. 86, 1315–1323. https://doi.org/10.1016/j.anbehav.2013.10.002

Sato, S., Tarumizu, K., Hatae, K., 1993. The influence of social factors on allogrooming in cows. Appl. Anim. Behav. Sci. 38, 235–244.

Schein, M.W., Fohrman, M.H., 1955. Social dominance relationships in a herd of dairy cattle. Br. J. Anim. Behav. 3, 45–55. https://doi.org/10.1016/S0950-5601(55)80012-3

Shimada, M., Sueur, C., 2014. The importance of social play network for infant or juvenile wild chimpanzees at Mahale Mountains National Park, Tanzania. Am. J. Primatol. n/a-n/a. https://doi.org/10.1002/ajp.22289

Snijders, L., Blumstein, D.T., Stanley, C.R., Franks, D.W., 2017. Animal Social Network Theory Can Help Wildlife Conservation. Trends Ecol. Evol. 32, 567–577. https://doi.org/10.1016/j.tree.2017.05.005

Sosa, S., 2016. The Influence of Gender, Age, Matriline and Hierarchical Rank on Individual Social Position, Role and Interactional Patterns in Macaca sylvanus at 'La Forêt des Singes': A Multilevel Social Network Approach. Front. Psychol. 7.

Sueur, C., Jacobs, A., Amblard, F., Petit, O., King, A.J., 2011a. How can social network analysis improve the study of primate behavior? Am. J. Primatol. 73, 703–719.

Sueur, C., King, A.J., Conradt, L., Kerth, G., Lusseau, D., Mettke-Hofmann, C., Schaffner, C.M., Williams, L., Zinner, D., Aureli, F., 2011b. Collective decision-making and fission–fusion dynamics: a conceptual framework. Oikos 120, 1608–1617. https://doi.org/10.1111/j.1600-0706.2011.19685.x

Sueur, C., Kuntz, C., Debergue, E., Keller, B., Robic, F., Siegwalt-Baudin, F., Richer, C., Ramos, A., Pelé, M., 2017. Leadership linked to group composition in Highland cattle (Bos taurus): Implications for livestock management. Appl. Anim. Behav. Sci. https://doi.org/10.1016/j.applanim.2017.09.014

Sueur, C., Pelé, M., 2015. Utilisation de l'analyse des réseaux sociaux dans la gestion des animaux maintenus en captivité, in: Analyse Des Réseaux Sociaux Appliquée à l'Ethologie et à l'Ecologie. Editions Matériologiques, pp. 445–468.

Syme, L.A., Syme, G., Waite, T., Pearson, A., 1975. Spatial distribution and social status in a small herd of dairy cows. Anim. Behav. 23, 609–614.

Takeda, K., Sato, S., Sugawara, K., 2003. Familiarity and group size affect emotional stress in Japanese Black heifers. Appl. Anim. Behav. Sci. 82, 1–11.

Tokuyama, N., Furuichi, T., 2017. Leadership of old females in collective departures in wild bonobos (Pan paniscus) at Wamba. Behav. Ecol. Sociobiol. 71, 55. https://doi.org/10.1007/s00265-017-2277-5

VILLARET, J.-C., BON, R., 1998. Sociality and relationships in Alpine ibex (Capra ibex).

Von Keyserlingk, M., Olenick, D., Weary, D., 2008. Acute behavioral effects of regrouping dairy cows. J. Dairy Sci. 91, 1011–1016.

Whitehead, H., 2009. SOCPROG programs: analysing animal social structures. Behav. Ecol. Sociobiol. 63, 765–778. https://doi.org/10.1007/s00265-008-0697-y

Wilson, A.D.M., Krause, S., Ramnarine, I.W., Borner, K.K., Clément, R.J.G., Kurvers, R.H.J.M., Krause, J., 2015. Social networks in changing environments. Behav. Ecol. Sociobiol. 1–13. https://doi.org/10.1007/s00265-015-1973-2

Winfield, C., Syme, G., Pearson, A., 1981. Effect of familiarity with each other and breed on the spatial behaviour of sheep in an open field. Appl. Anim. Ethol. 7, 67–75.

Wintz, M., Fabien, D., 2012. La perception des friches dans les Vosges du nord: entre nature abandonnée et nature «déjà là. Ann. Sci. Réserve Biosph. Transfront.

Wolf, M., Weissing, F.J., 2012. Animal personalities: consequences for ecology and evolution. Trends Ecol. Evol. 27, 452–461. https://doi.org/10.1016/j.tree.2012.05.001

Wrangham, R.W., 1980. An Ecological Model of Female-Bonded Primate Groups. Behaviour 75, 262–300.

Supplementary information for :

Impact of group management and transfer on individual sociality in Highland cattle (*Bos Taurus*)

Sebastian Sosa[1], Marie Pelé[2], Élise Debergue[3,] Cédric Kuntz[3], Blandine Keller[3], Florian Robic[3], Flora Siegwalt-Baudin[3], Camille Richer[3], Amandine Ramos[3], Cédric Sueur[3]

[1] Anthropology Department, Sun-Yat sen University, Guang Zhou, China

[2] Ethobiosciences, Research and Consultancy Agency in Animal Well-Being and Behaviour, Strasbourg, France

[3] Université de Strasbourg, CNRS, IPHC UMR 7178, F-67000 Strasbourg, France

Corresponding author: cedric.sueur@iphc.cnrs.fr, 0033(0)88107453, IPHC UMR 7178, 23 rue Becquerel F-67000 Strasbourg, France

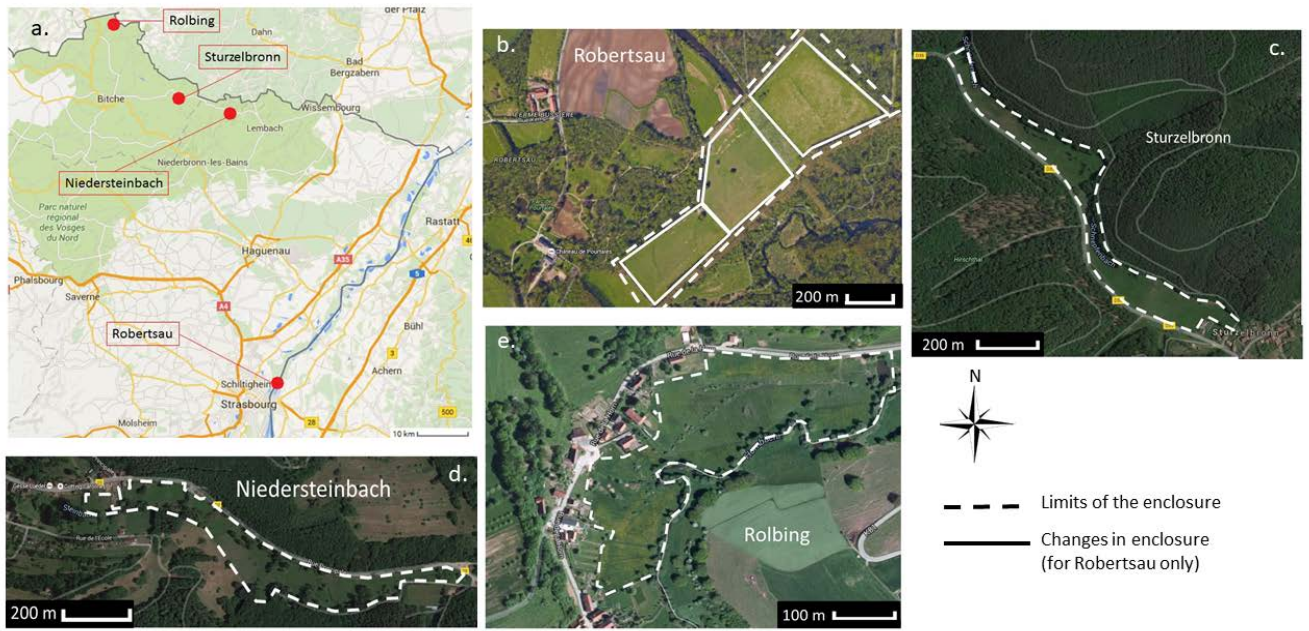

Figure S1: (a.) locations of the four different sites, (b.) Robertsau study site with the different enclosures (as full line), (c.) Sturzelbronn study site, (d.) Niedersteinbach study site and (e.) Rolbing study site.

Figure S2: Plots of residual normality distribution for the eigenvector centrality (a.), the strength of associations (b.), the difference of eigenvector centrality between two transfers (c.) and the difference of strength of associations between two transfers (d.).

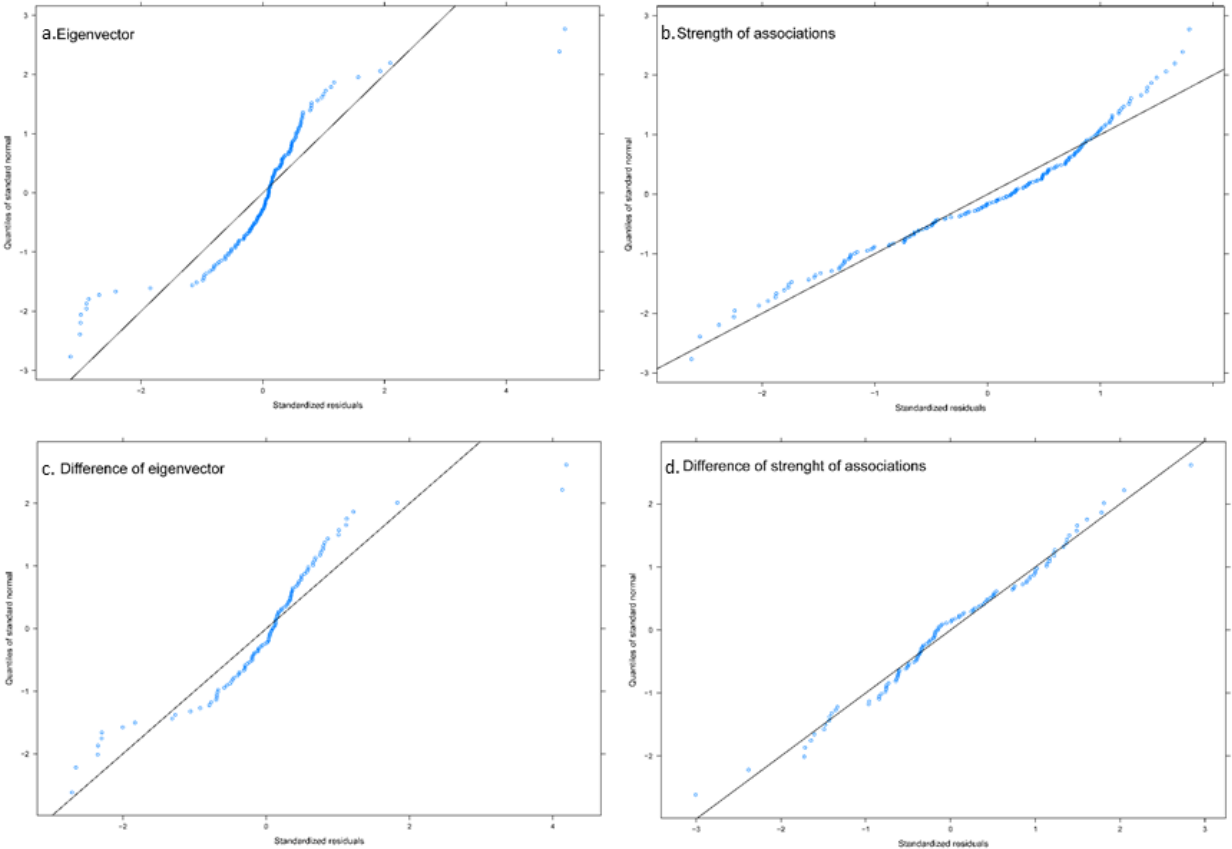

Table S1: Model selection table for the influence of socio-demographic factors on the eigenvector centrality. Models are ranked according to the best AIC. In bold, the models retained for the p-values average

| Model n° | (Int) | age | domin | famil | sex | df | logLik | AICc | ΔAIC | weight |
|---|---|---|---|---|---|---|---|---|---|---|
| **1** | **0.381** | | | | | **3** | **194.6** | **-383** | **0** | **0.736** |
| **3** | **0.408** | | **-0.05** | | | **4** | **194.4** | **-381** | **2.47** | **0.215** |
| **5** | **0.407** | | | **-0.03** | | **4** | **192.5** | **-377** | **6.28** | **0.032** |
| **7** | **0.433** | | **-0.05** | **-0.03** | | **5** | **192.3** | **-374** | **8.9** | **0.009** |
| **2** | **0.368** | **0.004** | | | | **4** | **190.8** | **-373** | **9.81** | **0.005** |
| 9 | 0.38 | | | | + | 5 | 190.4 | -371 | 12.57 | 0.001 |
| 4 | 0.398 | 0.002 | -0.04 | | | 5 | 189.7 | -369 | 14 | 0.001 |
| 11 | 0.388 | | -0.05 | | + | 6 | 190.5 | -369 | 14.51 | 0.001 |
| 6 | 0.399 | 0.005 | | -0.04 | | 5 | 189 | -368 | 15.5 | 0 |
| 13 | 0.383 | | | -0.03 | + | 6 | 188.4 | -364 | 18.79 | 0 |
| 8 | 0.424 | 0.003 | -0.04 | -0.03 | | 6 | 187.8 | -363 | 20.08 | 0 |
| 15 | 0.391 | | -0.06 | -0.03 | + | 7 | 188.6 | -363 | 20.59 | 0 |
| 10 | 0.368 | 0.005 | | | + | 6 | 186.9 | -361 | 21.8 | 0 |
| 12 | 0.38 | 0.002 | -0.05 | | + | 7 | 186 | -357 | 25.84 | 0 |
| 14 | 0.37 | 0.005 | | -0.04 | + | 7 | 185.2 | -356 | 27.44 | 0 |
| 16 | 0.382 | 0.003 | -0.05 | -0.04 | + | 8 | 184.2 | -352 | 31.56 | 0 |

Table S2: Model selection table for the influence of socio-demographic factors on the strength of associations. Models are ranked according to the best AIC. In bold, the models retained for the p-values average. Effect of sex is indicated by + because it is a factor (categorical).

| Model n° | (Int) | age | domin | famil | sex | df | logLik | AIC | ΔAIC | weight |
|---|---|---|---|---|---|---|---|---|---|---|
| **13** | **0.9447** | | | **0.2704** | **+** | **6** | **55.82** | **-99.6** | **0.0** | **0.707** |
| **14** | **0.9074** | **0.0147** | | **0.2465** | **+** | **7** | **54.98** | **-96** | **3.7** | **0.113** |
| **5** | **0.6884** | | | **0.1959** | | **4** | **51.77** | **-95.5** | **4.1** | **0.091** |
| **15** | **0.9531** | | **-0.0583** | **0.2663** | **+** | **7** | **54.36** | **-94.7** | **4.9** | **0.06** |
| **7** | **0.7191** | | **-0.0580** | **0.194** | | **5** | **50.28** | **-90.6** | **9.1** | **0.008** |
| **1** | **0.8599** | | | | | **3** | **48.23** | **-90.5** | **9.2** | **0.007** |
| **16** | **0.9119** | **0.0139** | **-0.0172** | **0.246** | **+** | **8** | **52.92** | **-89.8** | **9.8** | **0.005** |
| 6 | 0.6651 | 0.0124 | | 0.1749 | | 5 | 49.70 | -89.4 | 10.2 | 0.004 |
| 9 | 0.9718 | | | | + | 5 | 48.46 | -86.9 | 12.7 | 0.001 |
| 2 | 0.808 | 0.0155 | | | | 4 | 47.38 | -86.8 | 12.9 | 0.001 |
| 10 | 0.9228 | 0.0181 | | | + | 6 | 49.18 | -86.4 | 13.3 | 0.001 |
| 3 | 0.8929 | | -0.0658 | | | 4 | 46.91 | -85.8 | 13.8 | 0.001 |
| 8 | 0.6824 | 0.0113 | -0.0276 | 0.1755 | | 6 | 47.76 | -83.5 | 16.1 | 0 |
| 11 | 0.9818 | | -0.0736 | | + | 6 | 47.42 | -82.8 | 16.8 | 0 |
| 4 | 0.8254 | 0.0143 | -0.0265 | | | 5 | 45.44 | -80.9 | 18.8 | 0 |
| 12 | 0.9282 | 0.0171 | -0.0211 | | + | 7 | 47.19 | -80.4 | 19.3 | 0 |

Table S3: Model selection table for the influence of socio-demographic factors on the difference of eigenvector centralities after a transfer. Models are ranked according to the best AIC. In bold, the models retained for the p-values average

| Model n° | (Intrc) | age | domin | famil | nb.ind | df | logLik | AIC | ΔAIC | weight |
|---|---|---|---|---|---|---|---|---|---|---|
| **1** | **-0.0096** | | | | | **3** | **102.32** | **-198.6** | **0.0** | **0.854** |
| **2** | **-0.0400** | **0.0092** | | | | **4** | **100.56** | **-193.1** | **5.5** | **0.054** |
| **5** | **0.0073** | | | **-0.0185** | | **4** | **100.40** | **-192.8** | **5.9** | **0.046** |
| **3** | **-0.0090** | | **0.0171** | | | **4** | **100.17** | **-192.3** | **6.3** | **0.036** |
| 9 | -0.0107 | | | | -0.0022 | 4 | 97.62 | -187.2 | 11.4 | 0.003 |
| 6 | -0.0308 | 0.0091 | | -0.0099 | | 5 | 98.57 | -187.1 | 11.5 | 0.003 |
| 7 | 0.0217 | | 0.0278 | -0.0332 | | 5 | 98.43 | -186.9 | 11.8 | 0.002 |
| 4 | -0.0418 | 0.0096 | -0.0146 | | | 5 | 98.41 | -186.8 | 11.8 | 0.002 |
| 13 | 0.0155 | | | -0.0287 | -0.0024 | 5 | 95.78 | -181.6 | 17.1 | 0 |
| 8 | -0.0396 | 0.0096 | -0.0138 | -0.0023 | | 6 | 96.52 | -181 | 17.6 | 0 |
| 11 | -0.0102 | | 0.0111 | | -0.0021 | 5 | 95.43 | -180.9 | 17.8 | 0 |
| 10 | -0.0390 | 0.0088 | | | -0.0008 | 5 | 95.39 | -180.8 | 17.9 | 0 |
| 15 | 0.0275 | | 0.0238 | -0.0409 | -0.0023 | 6 | 93.77 | -175.5 | 23.1 | 0 |
| 14 | -0.0255 | 0.0086 | | -0.0143 | -0.0009 | 6 | 93.44 | -174.9 | 23.8 | 0 |
| 12 | -0.0408 | 0.0092 | -0.0155 | | -0.0008 | 6 | 93.25 | -174.5 | 24.2 | 0 |
| 16 | -0.0340 | 0.0090 | -0.0130 | -0.0070 | -0.0009 | 7 | 91.39 | -168.8 | 29.9 | 0 |

Table S4: Model selection table for the influence of socio-demographic factors on the difference of strength of associations after a transfer. Models are ranked according to the best AIC. In bold, the models retained for the p-values average

| Model n° | Int. | Age | Domin | Famil | Nb.ind | df | logLik | AIC | ΔAIC | weight |
|---|---|---|---|---|---|---|---|---|---|---|
| **7** | **-0.6639** | | **0.242** | **0.7524** | | **5** | **4.422** | **1.2** | **0.0** | **0.473** |
| **5** | **-0.7887** | | | **0.8802** | | **4** | **3.352** | **1.3** | **0.1** | **0.441** |
| **6** | **-0.8732** | **0.0203** | | **0.8992** | | **5** | **1.985** | **6** | **4.9** | **0.041** |
| **13** | **-0.7529** | | | **0.8354** | **-0.0103** | **5** | **1.101** | **7.8** | **6.6** | **0.017** |
| **15** | **-0.6394** | | **0.2252** | **0.7199** | **-0.0095** | **6** | **1.89** | **8.2** | **7.1** | **0.014** |
| **8** | **-0.7600** | **0.0150** | **0.1768** | **0.8009** | | **6** | **1.884** | **8.2** | **7.1** | **0.014** |
| 14 | -0.8281 | 0.0158 | | 0.8619 | -0.0076 | 6 | -1.279 | 14.6 | 13.4 | 0.001 |
| 16 | -0.7085 | 0.0102 | 0.1839 | 0.7581 | -0.0079 | 7 | -1.281 | 16.6 | 15.4 | 0 |
| 3 | 0.0318 | | 0.4848 | | | 4 | -7.173 | 22.3 | 21.2 | 0 |
| 11 | 0.0244 | | 0.4490 | | -0.0125 | 5 | -8.839 | 27.7 | 26.5 | 0 |
| 4 | 0.0211 | 0.0031 | 0.4744 | | | 5 | -10.745 | 31.5 | 30.3 | 0 |
| 1 | 0.0156 | | | | | 3 | -14.542 | 35.1 | 33.9 | 0 |
| 12 | 0.0370 | -0.0038 | 0.4599 | | -0.0130 | 6 | -12.375 | 36.8 | 35.6 | 0 |
| 9 | 0.0080 | | | | -0.0155 | 4 | -15.18 | 38.4 | 37.2 | 0 |
| 2 | -0.0375 | 0.0160 | | | | 4 | -17.066 | 42.1 | 41.0 | 0 |
| 10 | -0.0182 | 0.0081 | | | -0.0142 | 5 | -18.492 | 47 | 45.8 | 0 |